\documentclass[reprint,aps,pre,amsmath,amssymb,superscriptaddress,floatfix]{revtex4-2}
\usepackage{graphicx}
\usepackage{dcolumn}   
\usepackage{bm}         
\usepackage{xcolor}
\usepackage[dvipsnames]{xcolor}
\usepackage{tikz}
\usepackage[colorlinks=true,citecolor=black,linkcolor=black,urlcolor=blue]{hyperref}
\usepackage{subcaption}
\newif\ifshowcomments
\showcommentstrue      

\begin{document}
\raggedbottom

\title{Phase diagram of a semiflexible magnetic polymer in three dimensions}

\author{Linda Delimi}
\author{Arpan Dey}
\author{Jean-Charles Walter}
\email{jean-charles.walter@umontpellier.fr}
\affiliation{Laboratoire Charles Coulomb (L2C), Univ. Montpellier, CNRS, Montpellier, France}

\date{\today} 

\begin{abstract}
We study a semiflexible magnetic polymer in which each monomer carries a two-state spin, giving rise to coupled conformational, stiffness, and magnetic degrees of freedom. The model provides a minimal representation of nucleoprotein complexes {\it in vivo}, in which DNA bending rigidity and bridging proteins must be simultaneously accommodated. We employ Monte Carlo simulations on a face-centered-cubic (FCC) lattice, complemented by a variational free-energy upper-bound approach, to characterize the zero-field phase diagram as a function of bending stiffness and temperature. At low stiffness, we identify a first-order transition combining coil-to-globule collapse with a ferromagnetic-to-paramagnetic transition. At lower temperatures, an additional first-order transition separates the disordered globule from an ordered crystal characterized by nematic-like orientational order of the bond tangents. At high stiffness, the globule disappears, resulting in a direct first-order coil-to-crystal transition. These results demonstrate how chain stiffness and magnetic degrees of freedom couple to polymer conformation and reorganize the zero-field phase diagram. Our model connects the two limiting cases of flexible magnetic polymers and semiflexible homopolymers, the latter corresponding to the strong-field limit in which all spins are aligned.
\end{abstract}

\maketitle

\section{Introduction}

Biological polymers such as DNA are subject, {\it in vivo}, to various physical constraints that allow them to fit within a crowded cellular environment while remaining accessible to various biological processes involved namely in replication, transcription, and segregation. Their conformations are governed not only by entropy and excluded-volume interactions, but also by local bending stiffness and interactions mediated by bound proteins. A useful minimal question is therefore how bending stiffness and bridging interactions jointly reshape the equilibrium conformations of a polymer. This question is addressed in the present paper using a semiflexible magnetic polymer model, which provides a controlled framework for investigating the interplay between chain stiffness and bridging interactions.

Each DNA binding site can be either vacant or occupied by a bridging protein. This binary variable can be represented by an Ising-like spin. A magnetic polymer is a polymer whose monomers carry an internal degree of freedom, such as an Ising spin, that couples to spatial contacts~\cite{Garel1999PhasePolymers,Foster2021CriticalDimensions}. In the minimal lattice formulation, short-range contacts between monomers contribute an energy that depends on their spin alignment. As a result, magnetic ordering and conformational collapse become intertwined: like-spin correlations favour energetically favourable contacts and promote compaction, while compaction increases the number of possible contacts and, in turn, reinforces magnetic correlations. 

Such coupled models have been proposed as minimal descriptions of chromosome-folding scenarios in eukaryotes, in which compaction is linked to the spatial organization of biochemical marks~\cite{Jost2014,Coli2019,Gurgo2024}.

Another relevant biological {\it in vivo} system is the nucleoprotein complex ParB{\it S}, which is involved in bacterial DNA segregation~\cite{lorentz2025future}. The ParB proteins bind specifically at the {\it parS} sequence of the replicon, triggering the clustering of hundreds of ParB in a liquid-like droplet~\cite{Delimi2025,Guilhas2020,Revoil2026}. Interestingly this condensation is triggered by the use of CTP~\cite{Jalal2020,osorio2021thectp,Walter2020,Delimi2025} and this high concentration of ParB catalyzes the ATP hydrolysis of ParA that performs the actual segregation~\cite{Walter2017}.

Due to recent high-throughput sequencing methods like ChIP-sequencing, the high-resolution binding profile of ParB has been uncovered, namely on the F-plasmid~\cite{Sanchez2015,debaugny2018conserved}. The scale of the ParB footprint is 16 bp, i.e., much smaller than the persistence length of DNA, which is approximately 150 bp. This separation of length scales raises the question of how the phase behaviour of the magnetic polymer is modified by chain stiffness. Extending the phase diagram of the magnetic polymer along the persistence-length axis is therefore a natural and necessary step toward connecting the minimal model to the physical properties of DNA.

Garel {\it et al.} introduced the flexible magnetic polymer and established its phase diagram~\cite{Garel1999PhasePolymers}. More recently, Foster {\it et al.} revisited this model, confirmed the critical properties of the coupled transitions, and emphasized the role of dimensionality and universality~\cite{Foster2021CriticalDimensions}. The lattice-gas version of the magnetic polymer, known as the spreading-and-bridging model, has also been applied to the formation of ParB{\it S} complexes~\cite{broedersz2014condensation,David2020}. However, both studies failed to reproduce the high-resolution ChIP-sequencing profiles~\cite{Sanchez2015}. None of these studies systematically investigated the effect of the polymer persistence length on the transitions.
To date, the only model that has successfully described the high-resolution ChIP-seq data is the Stochastic Binding model. However, this is a phenomenological model that assumes a pre-existing ParB droplet, with a polymer attached to and fluctuating within it. It nevertheless demonstrated the necessity of a long-range power-law decay of ParB~\cite{Sanchez2015,debaugny2018conserved,walter2021supercoiled,Walter2018}.
Thus, no bottom-up model describing the formation of nucleoprotein condensates at the scale of the persistence length is currently available.

However, several useful limits have been established in the literature. First, for the flexible magnetic polymer in zero field, the system exhibits a first-order coil-to-globule transition~\cite{Garel1999PhasePolymers,Foster2021CriticalDimensions}. This regime is disrupted by an external field of a few times the thermal energy, although the precise location of this tricritical point is difficult to estimate~\cite{Garel1999PhasePolymers,Foster2021CriticalDimensions}. Beyond this critical field, the spins become sufficiently aligned for the transition to enter the $\theta$-like universality family. It provides a paradigmatic setting for the coil--globule transition and its scaling description~\cite{degennes_scaling,flory_polymer}. Under strong coupling (or poor-solvent conditions), attractive interactions between monomers compete with configurational entropy, driving the $\theta$-like collapse from a swollen coil to a compact, disordered globule.

Another useful limit is thus this large-field limit, where all spins are aligned and the model reduces to a homopolymer, which has been studied with a finite bending stiffness, making local curvature energetically costly. This interplay has been clarified in numerous studies of interacting semiflexible homopolymers. Bascle \emph{et al.} studied the dense semiflexible homopolymer using a mean-field representation of Hamiltonian paths, introducing an anisotropic mean-field treatment to describe the orientationally ordered low-temperature phase~\cite{Bascle1993}. Doniach \emph{et al.} developed a mean-field description and showed that bending stiffness reorganizes the low-temperature phase diagram, with a first-order transition between ``crystal-like'' compact phases and ``disordered'' compact states~\cite{Doniach1996PhaseFolding}. Subsequent work by Doye \emph{et al.} showed that increasing stiffness can shrink and eventually eliminate the intermediate disordered globular regime, leading, for sufficiently stiff chains, to a direct first-order transition between a compact ordered phase and a swollen coil~\cite{Doye1997TheHomopolymers}. Together, these results establish that bending stiffness can qualitatively modify the phase behavior of semiflexible polymers. More recent off-lattice studies of semiflexible homopolymers~\cite{Arcangeli2024PhaseBehaviorSemiflexiblePoorSolvent} have shown that the competition between attractive interactions and bending stiffness can produce an even richer structural phase diagram, including helical, rod-like, and sheet-like folded states.

Here, we bridge these limiting cases by studying the semiflexible magnetic polymer at zero field on a three-dimensional face-centered-cubic (FCC) lattice. The zero-field limit is the most interesting as it leaves the spins free to fluctuate, and thus enriches the phenomenology compared to the large-field regime, which falls into the $\theta$-like universality class.

The full model is controlled by the bending stiffness, the external magnetic field, and the temperature. Setting the external field to zero, we therefore focus on establishing the stiffness--temperature phase diagram. Each monomer of the chain carries an Ising spin that interacts with neighboring spins through spatial contacts and couples to the external magnetic field, while local bends are penalized by a stiffness term.

We first characterize the ground state which displays a dense folded structure characterized by straight stretches of polymer, with curvature localized at the surface. This structure starts from a quasi-isotropic shape at low stiffness and elongates progressively as the stiffness increases, in order to reduce the number of bends, which become increasingly costly. Knowledge of the ground state is useful both for the numerical simulations, to set the initial state, and for the analytics, to construct the thermodynamic-limit phase diagram from upper bounds of the free energy.

Secondly, we use Monte Carlo simulations to characterize the conformational, magnetic, and orientational transitions and map the resulting zero-field phase behavior in the stiffness--temperature plane. We observe clear signatures of first-order transitions along each transition line, together with long-lived metastable states between the two coexisting thermodynamic states.

Finally, we also investigate the model analytically. At zero field, the spins are free to fluctuate, while the bending stiffness couples directly to the local geometry, so that polymer conformation, curvature, spatial contacts, and magnetic ordering are coupled simultaneously. Rather than introducing several coupled auxiliary fields through a standard Landau or Hubbard--Stratonovich treatment, we adopt a phase-wise variational free-energy upper-bound approach inspired by the construction of Garel {\it et al.}~\cite{Garel1999PhasePolymers}: for the coil and globule phases, we extend the magnetic-polymer bounds by including an FCC bending contribution, while the compact ordered crystal phase is represented by a simple geometrically motivated ansatz. The resulting analytical phase diagram is obtained by comparing these free-energy bounds in the thermodynamic limit and provides a mean-field-level interpretation of the numerical results.

Both numerics and anlaytics lead to the same conclusion: At low stiffness, we identify a first-order transition combining coil-to-globule collapse with a ferromagnetic-to-paramagnetic transition. At lower temperatures, an additional first-order transition separates the disordered globule from an ordered crystal characterized by nematic-like orientational order of the bond tangents. At high stiffness, the globule disappears, resulting in a direct first-order coil-to-crystal transition. These results demonstrate how chain stiffness and magnetic degrees of freedom couple to polymer conformation and reorganize the zero-field phase diagram. 

\section{The Model}

\begin{figure}[!h]
    \centering

\begin{tikzpicture}[
    x=0.02cm,
    y=-0.02cm,
    >=latex,
    line cap=round,
    line join=round,
    font=\Large,
    monomere/.style={circle, minimum size=4.6mm, inner sep=0pt, draw=none}
]

\definecolor{chaine}{RGB}{30,30,30}
\definecolor{spinplus}{RGB}{69,151,247}
\definecolor{spinminus}{RGB}{212,53,46}
\definecolor{interaction}{RGB}{242,174,64}
\definecolor{rigidite}{RGB}{158,47,203}

\path[use as bounding box] (28,5) rectangle (372,325);

\draw[chaine, line width=4.2pt]
    plot[smooth, tension=0.55] coordinates {
        (63,71)
        (84,76)
        (113,88)
        (140,104)
        (165,121)
        (192,145)
        (211,161)
        (239,173)
        (272,173)
        (302,158)
        (327,138)
        (342,114)
        (343,73)
        (325,57)
        (300,45)
        (269,46)
        (233,64)
        (207,90)
        (191,115)
        (172,159)
        (173,189)
        (181,216)
        (187,241)
        (177,276)
        (153,302)
        (129,308)
        (101,301)
        (78,284)
        (73,253)
        (89,232)
        (113,222)
        (136,214)
        (145,190)
        (130,164)
        (106,155)
        (73,146)
        (53,144)
    };

\foreach \p in {
    (84,76),(140,104),(165,121),(211,161),(239,173),
    (302,158),(327,138),(269,46),(233,64),(172,159),
    (187,241),(177,276),(101,301),(73,253),(89,232),
    (136,214),(106,155),(73,146)
}{
    \node[monomere, fill=spinminus] at \p {};
}

\foreach \p in {
    (113,88),(192,145),(272,173),(342,114),(343,73),
    (325,57),(300,45),(207,90),(191,115),(173,189),
    (181,216),(153,302),(129,308),(78,284),(113,222),
    (145,190),(130,164)
}{
    \node[monomere, fill=spinplus] at \p {};
}

\draw[interaction, <->, line width=1.4pt]
    (146,189) -- (172,189);
\node[interaction, font=\bfseries] at (159,214) {$-J$};
\node[spinplus, anchor=east] at (137,190)
    {$\sigma_i=+1$};

\node[spinplus, anchor=west] at (181,190)
    {$\sigma_j=+1$};
\draw[->, line width=1.1pt] (217,58) -- (257,35)
    node[pos=0.25, above left=-1pt] {$\vec t_i$};

\draw[->, line width=1.1pt] (276,34) -- (316,34)
    node[midway, above=2pt] {$\vec t_{i+1}$};

\node[rigidite, font=\bfseries] at (263,14) {$\kappa$};
\end{tikzpicture}
    \caption{\textbf{Schematic of the semiflexible magnetic polymer.} Each monomer carries a spin $\sigma_i=\pm1$. Spatially neighbouring spins interact with coupling $J$ and couple to an external field $h$, while consecutive bond tangents $\hat{\mathbf t}_i$ and $\hat{\mathbf t}_{i+1}$ incur a bending penalty controlled by the stiffness $\kappa$.}
    \label{fig:model}
\end{figure}
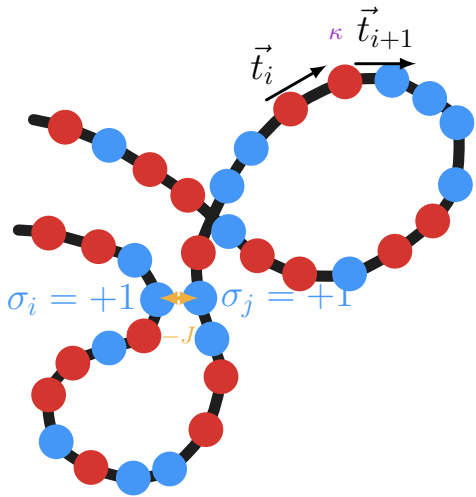

In this section, we describe the Hamiltonian of the semiflexible magnetic polymer. We consider a self-avoiding polymer chain composed of $N$ monomers, with fixed bond length $a$. In the following, all lengths are expressed in units of $a$, such that $a=1$. The position of each monomer $i$ is defined by a three-dimensional vector $\mathbf{r}_i$, and each monomer carries an internal degree of freedom $\sigma_i=\pm 1$.

First, we consider the nearest-neighbour spin--spin coupling term:
\begin{equation}
\mathcal{H}_c = - J\sum_{\langle i,j\rangle} \sigma_i \sigma_j\,,
\end{equation}
where $\langle i,j\rangle$ denotes pairs of monomers that are nearest neighbours on the lattice, whether or not they are consecutive along the polymer chain. The coupling constant $J$ sets the strength of the interaction between the spins. We take $J>0$, corresponding to ferromagnetic spin coupling.

The relative populations of spins pointing up and down are controlled by the linear coupling to the external magnetic field:
\begin{equation}
\mathcal{H}_f = - h \sum_{i=1}^{N} \sigma_i\,,
\end{equation}
where $h$ denotes the external magnetic field. For $h>0$, the field favours spins with $\sigma_i=+1$, while $h<0$ favours spins with $\sigma_i=-1$.

Finally, we energetically penalize local curvature through a discrete wormlike-chain Hamiltonian:
\begin{eqnarray}
\mathcal{H}_b &=& \kappa \sum_{i=1} ^ {N-2} \left(1-\hat{\mathbf{t}}_i\cdot \hat{\mathbf{t}}_{i+1}\right),\\
                    &=&\kappa \sum_{i=1} ^ {N-2} \left(1-\cos\theta _{i}\right)\,,
\end{eqnarray}
where $\hat{\mathbf{t}}_i = (\mathbf{r}_{i+1}-\mathbf{r}_i)/|\mathbf{r}_{i+1}-\mathbf{r}_i|$ is the unit tangent vector along bond $i$, $\theta_i$ is the angle between two consecutive bonds, and $\kappa$ is the bending stiffness.

The overall Hamiltonian is therefore given by the sum of these three contributions:
\begin{equation}
\mathcal{H}=\mathcal{H}_c+\mathcal{H}_f+\mathcal{H}_b\,.
\end{equation}

We focus here on the zero-field case, $h=0$, for which $\mathcal{H}_f=0$, while the spin degrees of freedom remain free to fluctuate. In the large-field limit, the spins become polarized, and the model reduces to a semiflexible homopolymer. Therefore, the $h\to0$ limit defines the new regime of the phase diagram. The field strength at which the transition between magnetic-like and homopolymer-like behaviour occurs is notoriously difficult to estimate even in the flexible case~\cite{Garel1999PhasePolymers,Foster2021CriticalDimensions}, although it is expected to be of the order of a few times the thermal energy $kT$. By analogy, the semiflexible magnetic polymer is expected to exhibit a non-trivial phase diagram in the limit $h\to0$, which motivates our approach.

We also define the reduced variables
\begin{equation}
    \mathcal{T}=\frac{T}{J},\qquad
    \mathcal{K}=\frac{\kappa}{J},
\end{equation}
where $T$ is the temperature of the system.

In the following, we embed the polymer in a three-dimensional face-centered-cubic (FCC) lattice of linear size $L_{\rm box}$ with hard-wall boundary conditions.

\section{Ground state of the semiflexible magnetic polymer}
\label{subsec:low_energy_stiff}

\paragraph{Energy minimization of dense folded structures}

To characterize the effect of stiffness on the low-energy conformations of the semiflexible magnetic polymer, we consider a family of dense folded structures constructed on the FCC lattice~\cite{Arcangeli2024PhaseBehaviorSemiflexiblePoorSolvent,Fisher2013FourDimensional,Wolf1999Biocrystallization,FrenkielKrispin2004Nucleoid,Metzler2004Helical,Petrov2008Packaging,Balhorn2007Protamine,Vertchik2025Helical}. Related compact, folded, helical, and toroidal morphologies have been reported in simulations and in experiments on different polymeric and DNA-condensed systems. Each candidate consists of a dense packing of straight segments traversed back and forth and connected by changes in direction, and is characterized by its three linear dimensions. At $h=0$ and $T=0$, the ferromagnetic ground state has all spins aligned. Since consecutive monomers are always nearest neighbours on the lattice, their spin contribution is independent of the polymer conformation and can be subtracted as a common additive constant. For fixed $N$, $\kappa$, and $J$, we vary the three linear dimensions of the folded structure and retain the configuration of minimum energy. The resulting ground-state energy is:

\begin{equation}
E_0=\min(-Jn_c+\kappa n_b)\,,
\label{eq:low_energy_folded}
\end{equation}

where $n_c$ denotes the number of nonbonded nearest-neighbor contacts 

\begin{equation}
n_c
=
\sum_{\substack{1\leq i<j\leq N\\ |i-j|>1}}
\Delta_{ij},
\label{eq:number_contacts}
\end{equation}

where $\Delta_{ij}=1$ if monomers $i$ and $j$ occupy nearest-neighbor lattice sites and $\Delta_{ij}=0$ otherwise, and $n_b$ measures the geometric cost associated with local changes in direction of the tangent  between two consecutive bonds

\begin{equation}
n_b =
\sum_{i=1}^{N-2}
\left(
1-\hat{\mathbf t}_i\cdot\hat{\mathbf t}_{i+1}
\right)\,.
\label{eq:bending_sum_groundstate}
\end{equation}

At zero temperature, the selected geometry is determined by the competition between maximizing favorable nonbonded contacts and minimizing the bending contribution. Longer straight segments reduce $n_b$ by decreasing the number of changes in direction, but this also exposes more monomers at the surface, where the number of favorable contacts is reduced. The ground-state structure therefore reflects a balance between contact gain and bending cost.

\paragraph{Aspect-ratio of the dense folded structure: competition between surface contacts and bending}

For each minimum-energy configuration, we denote its three linear dimensions, ordered by magnitude, as

\begin{equation}
a^\star
\leq
b^\star
\leq
c^\star.
\end{equation}

We then define the aspect ratio

\begin{equation}
\mathcal A^\star
=\frac{c^\star}
{\sqrt{a^\star b^\star}},
\label{eq:aspect_ratio_groundstate}
\end{equation}

which characterizes the elongation of the selected structure.

Increasing $\mathcal A^\star$ allows the straight segments to become longer and reduces the number of turns. It therefore decreases the bending contribution $n_b$. This reduction, however, comes at a cost: at fixed monomer number, a more elongated geometry has a larger surface area and loses more contacts relative to the bulk. The optimal finite-size geometry therefore reflects a competition between the loss of surface contacts and the bending cost associated with the turns.

\begin{figure}[htbp]
\centering
\includegraphics[width=0.5\textwidth]{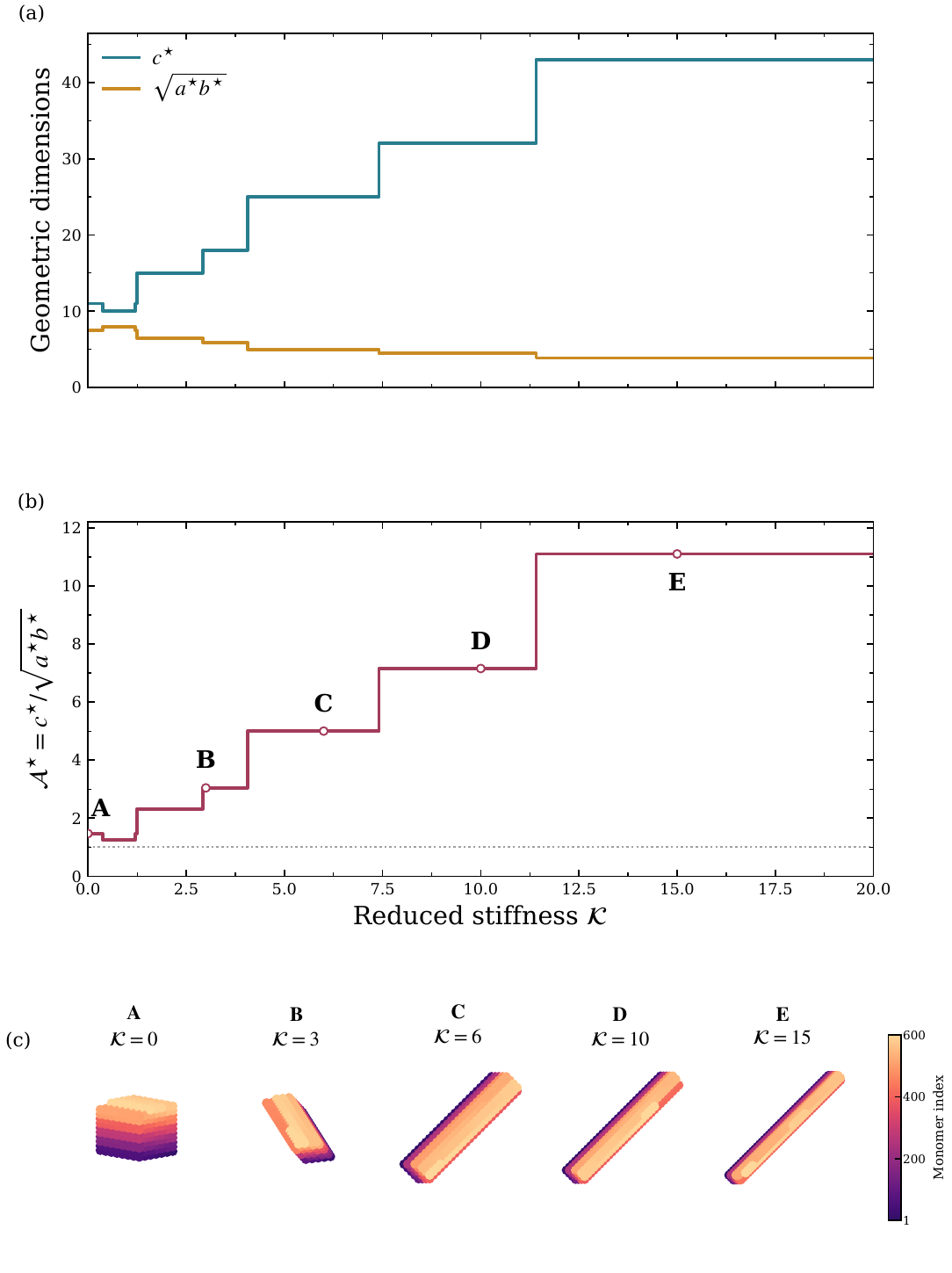}
\caption{\textbf{Aspect ratio $\mathcal A^\star
=c^\star/\sqrt{a^\star b^\star}$ and ordered dimensions of the minimum-energy dense folded structure as a function of the reduced stiffness $\kappa/J$, for $N=400$.}
\textbf{(a)} Aspect ratio of the selected configuration.
\textbf{(b)} Longitudinal dimension $c^\star$ and transverse dimension $\sqrt{a^\star b^\star}$. The stepwise variations correspond to discrete changes in the optimal geometry on the lattice.
(c) (A–E) Snapshots of the ground state at different values of $\mathcal{K}$. The corresponding letters in (B) indicate their precise locations in the phase diagram.
}  
\label{fig:groundstate_geometry}
\end{figure}

Figure~\ref{fig:groundstate_geometry}(a) shows that increasing stiffness selects progressively more elongated structures. The longitudinal dimension $c^\star$ increases, whereas the transverse dimension $\sqrt{a^\star b^\star}$ decreases. The aspect ratio $\mathcal A^\star$ therefore increases in a stepwise manner.

The plateaus observed in Fig.~\ref{fig:groundstate_geometry}(b) correspond to successive selections of distinct discrete geometries on the FCC lattice. To illustrate the associated structural rearrangements, we select five representative configurations, labeled A to E, from different plateaus of $\mathcal A^\star$, as shown in Fig.~\ref{fig:groundstate_geometry}(c).

This evolution directly reflects the energetic competition described above. Elongation reduces the number of changes in direction and hence the bending cost. Conversely, the surface area of the structure increases and the number of contacts decreases. Each step corresponds to the selection of a new geometry providing a more favorable balance between these two contributions.

This competition can be summarized by separating the extensive bulk energy from the leading finite-size corrections. For a dense structure whose aspect ratio remains finite, the bulk contact contribution scales as $N$, while the loss of surface contacts and the bending contribution both scale as $N^{2/3}$. We therefore write

\begin{equation}
\begin{aligned}
E_0(N,\mathcal A)
={}&-J n_c^{\mathrm{bulk}}N\\
&+\left[
J\lambda_{\mathrm{surf}}(\mathcal A)
+\kappa\lambda_{\mathrm{bend}}(\mathcal A)
\right]N^{2/3}
+\cdots ,
\end{aligned}
\label{eq:bulk_surface_bending}
\end{equation}

where $n_c^{\mathrm{bulk}}$ is the number of nonbonded bulk contacts per monomer, while $\lambda_{\mathrm{surf}}(\mathcal A)$ and $\lambda_{\mathrm{bend}}(\mathcal A)$ are dimensionless geometric coefficients describing the loss of surface contacts and the bending contribution, respectively.

\paragraph{Consequence in the thermodynamic limit}

The scaling form above makes the thermodynamic limit immediate. For a dense structure of finite aspect ratio, $E_0^{\mathrm{bulk}}\sim -N$, whereas the surface and bending corrections scale as $N^{2/3}$. The ground-state energy per monomer therefore reads

\begin{equation}
\frac{E_0}{N}
=
-Jn_c^{\mathrm{bulk}}
+
\mathcal O
\left(
N^{-1/3}
\right).
\label{eq:thermodynamic_limit_folded}
\end{equation}

Thus, the surface and bending contributions become negligible per monomer compared with the bulk contribution as $N\rightarrow\infty$. In this limit, the energy of compact structures is therefore dominated by their bulk contribution.

This property motivates the study of this simple geometric family. It allows us to isolate the mechanism by which stiffness modifies the shape at finite size while retaining the extensive contribution characteristic of a dense structure. Other compact morphologies may exhibit the same bulk behaviour while differing in their surface and bending contributions. They therefore provide natural extensions of this analysis.

\section{Numerical results}
\subsection{Observables}

The characterization of the phases relies primarily on three observables: the contact density \(\rho_c\), the magnitude of the magnetization \(|m|\), and the nematic order parameter \(S\). They respectively describe the compaction of the chain, magnetic order, and orientational order. The radius of gyration \(R_g\) complements this description by characterizing the spatial extent of the chain.

The chain compaction is characterized by the density of nonbonded contacts.
\begin{equation}
\rho_c
=
\frac{n_c}{N},
\label{eq:densite_contacts_sfmp}
\end{equation}
where \(n_c\) is the number of pairs of nonconsecutive monomers occupying nearest-neighbor sites on the lattice, as defined in Eq.~\eqref{eq:number_contacts}.

A high value of \(\rho_c\) characterizes a compact chain. The radius of gyration \(R_g\) provides a complementary measure of the spatial extent of the chain and is defined by
\begin{equation}
R_g^2
=
\frac{1}{N}
\sum_{i=1}^{N}
\left|
\mathbf r_i-\mathbf r_{\mathrm{cm}}
\right|^2,
\qquad
\mathbf r_{\mathrm{cm}}
=
\frac{1}{N}
\sum_{i=1}^{N}\mathbf r_i,
\label{eq:radius_gyration}
\end{equation}
where \(\mathbf r_{\mathrm{cm}}\) is the center-of-mass position of the chain. An extended chain typically has a large radius of gyration, whereas a compact chain has a smaller radius of gyration.
The contact density is used as the main indicator of compaction throughout the following, as it remains directly related to the number of contacts even for compact and elongated structures.

The magnetic order of the internal variables is quantified by the magnetization per monomer.
\begin{equation}
m
=
\frac{1}{N}
\sum_{i=1}^{N}\sigma_i.
\label{eq:aimantation_sfmp}
\end{equation}

At zero field, the two ferromagnetic orientations \(m>0\) and \(m<0\) are equivalent by symmetry. We therefore use \(\langle |m| \rangle\) as the magnetic order parameter.
A low value of \(\langle|m|\rangle\) characterizes the paramagnetic regime, whereas a high value characterizes the ferromagnetic regime.

The orientational organization of the chain is characterized by the nematic order tensor~\cite{DoiEdwards1986}.

\begin{equation}
Q_{\alpha\beta}
=
\frac{1}{N-1}
\sum_{i=1}^{N-1}
\left(
\frac{3}{2}
t_{i,\alpha}t_{i,\beta}
-
\frac{1}{2}\delta_{\alpha\beta}
\right),
\qquad
\alpha,\beta\in\{x,y,z\},
\label{eq:tenseur_nematique_revue}
\end{equation}
where \(t_{i,\alpha}\) denotes the component along the \(\alpha\) axis of the unit tangent vector \(\mathbf t_i\).
The nematic order parameter \(S\) is defined as the largest eigenvalue of this tensor,
\begin{equation}
    S=\lambda_{\max}(Q).
    \label{eq:ordre_nematique_revue}
\end{equation}
An isotropic distribution of orientations leads to a low value of \(S\), whereas preferential alignment of the segments along a common axis leads to a high value of \(S\). In the limit of perfect alignment, \(S=1\). The tensor \(Q\) is invariant under the transformation \(\mathbf t_i\rightarrow-\mathbf t_i\). Parallel and antiparallel segments therefore contribute equally to the nematic order.
The parameter \(S\) characterizes the orientation of the chain independently of its compaction. We therefore interpret it jointly with \(\rho_c\): a high value of \(S\) associated with a high contact density characterizes the oriented compact phase.

\subsection{Phase diagram of the semiflexible magnetic polymer}
\label{subsec:phase_diagram_sfmp}

\subsubsection{Phase diagram for \(N=600\)}
\begin{figure}[]
    \centering
    \includegraphics[width=\linewidth]{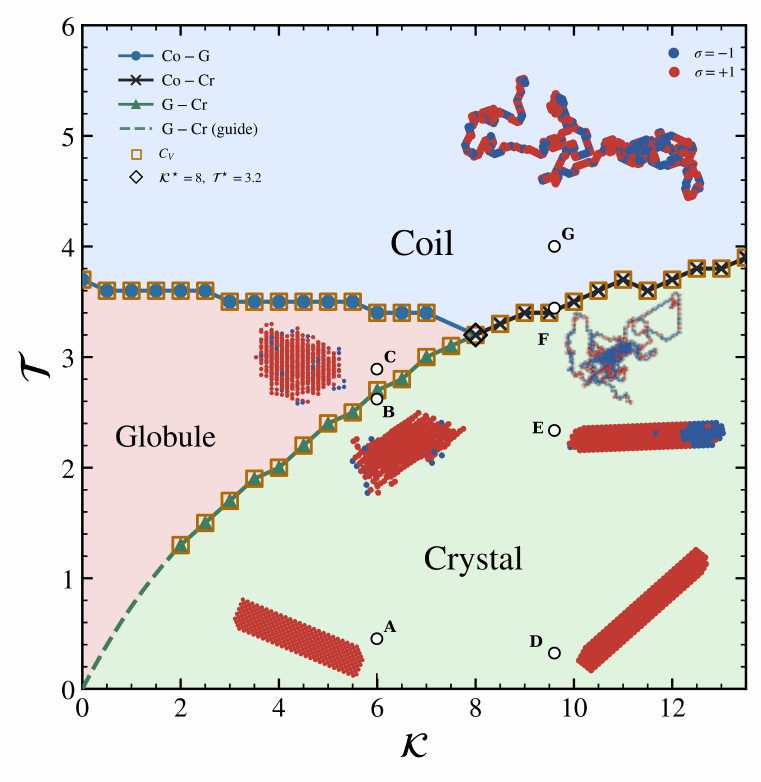}
    \caption{
    \textbf{Phase diagram of the semiflexible magnetic polymer at zero field for \(N=600\).} The coil, globule, and crystal regimes are indicated in blue, pink, and green, respectively. Blue circles, green triangles, and black crosses correspond to the Co--G, G--Cr, and Co--Cr boundaries, respectively; open orange squares indicate the resolved peaks of \(C_V\). The dashed green line represents the indicative continuation of the crystalline boundary at low stiffness. The open diamond marks the triple point \((\mathcal K^\star,\mathcal T^\star)\). Configurations A--G illustrate representative conformations at the positions indicated by the open circles; monomers are colored according to their magnetic state, \(\sigma=-1\) in blue and \(\sigma=+1\) in red.
    }
    \label{fig:sfmp_phase_diagram}
\end{figure}

We characterize the organization of the semiflexible magnetic polymer at zero field in the plane of the reduced parameters \((\mathcal K,\mathcal T)\). For each value of \(\mathcal K\), the simulations follow the heating protocol described in
Appendix~\ref{app:simulation_protocol}, initialized from the low-energy compact configuration obtained at \(\mathcal T=0\). The resulting diagram for a chain of \(N=600\) monomers (Fig.~\ref{fig:sfmp_phase_diagram}) distinguishes three regimes: the paramagnetic coil (Co), the disordered globule (G), and the crystalline phase (Cr).
The phase boundaries are located from the peaks of \(C_V\) and identified based on the maxima of the associated susceptibilities.
The Co--G transition is identified by the coincidence of a peak in \(C_V\) with the maximum of \(\chi_{\rho_c}\). It is resolved for \(2\lesssim\mathcal K\lesssim8\). For \(\mathcal K\gtrsim8\), the maxima of \(\chi_{\rho_c}\) and \(\chi_S\) cluster around the same peak in \(C_V\). Compaction and the establishment of nematic order then occur at the same temperature, defining a direct Co--Cr transition.
The G--Cr transition is identified by the peak in \(C_V\) associated with the maximum of \(\chi_S\). For \(2\lesssim\mathcal K\lesssim8\), it appears at a lower temperature than the Co--G transition, such that two distinct peaks in \(C_V\) are resolved in this region. For \(0<\mathcal K<2\), the G--Cr transition cannot be located separately from the \(C_V\) curves. The continuation of this boundary toward low stiffness is therefore indicated by a dashed line as a visual guide between \((\mathcal K,\mathcal T)=(0,0)\) and the first directly resolved points around \(\mathcal K\simeq2\).

\subsubsection{Coil--Globule transition line}
We characterize the Co--G transition through an analysis of the distributions, followed by a finite-size study at \(\mathcal K=6\).

\begin{figure*}[t]
    \centering

    \begin{subfigure}[t]{0.32\textwidth}
        \centering
        \includegraphics[width=\linewidth]{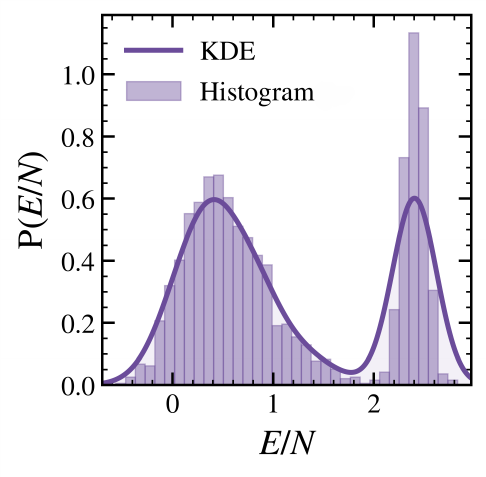}
        \caption{}
        \label{fig:cg_hist_energy_k6}
    \end{subfigure}
    \hfill
    \begin{subfigure}[t]{0.32\textwidth}
        \centering
        \includegraphics[width=\linewidth]{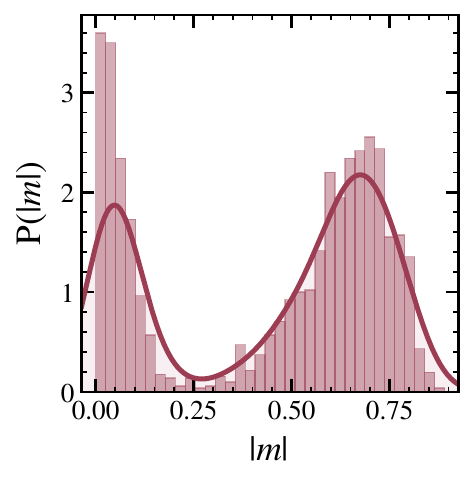}
        \caption{}
        \label{fig:cg_hist_m_k6}
    \end{subfigure}
    \hfill
    \begin{subfigure}[t]{0.32\textwidth}
        \centering
        \includegraphics[width=\linewidth]{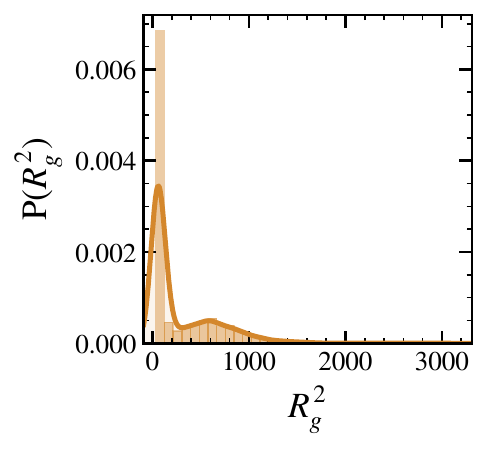}
        \caption{}
        \label{fig:cg_hist_Rg2_k6}
    \end{subfigure}

    \caption{
\textbf{Distributions of the observables in the vicinity of the Co--G transition.}
Distributions of the energy per monomer $E/N$ \textbf{(a)}, the absolute magnetization $|m|$ \textbf{(b)}, and the squared radius of gyration $R_g^2$ \textbf{(c)}, for $N=600$ and $\mathcal T=3.375$ at $\mathcal K=6$, approaching the critical value from below. The normalized histograms are complemented by Gaussian kernel density estimates (KDE).
    }
    \label{fig:cg_distributions}
\end{figure*}

At \(\mathcal K=6\), the transition is located from the coincidence of the peak of \(C_V\) with the maximum of \(\chi_{\rho_c}\). For \(N=600\), the average of the pseudocritical temperatures estimated over the replicas is \(\mathcal T_{\mathrm{CG}}=3.43\pm0.09\), with the uncertainty corresponding to their dispersion.
At \(\mathcal T=3.375\), the distributions of \(E/N\) and \(|m|\) exhibit two distinct populations (Fig.~\ref{fig:cg_distributions}). The distribution of \(R_g^2\) is strongly asymmetric and accompanies the change in polymer compaction. Taken together, these distributions highlight the coexistence of configurations with distinct energetic, magnetic, and structural properties in the vicinity of the Co--G transition.

\begin{figure*}[t]
    \centering

    \begin{subfigure}[t]{0.32\textwidth}
        \centering
        \includegraphics[width=\linewidth]{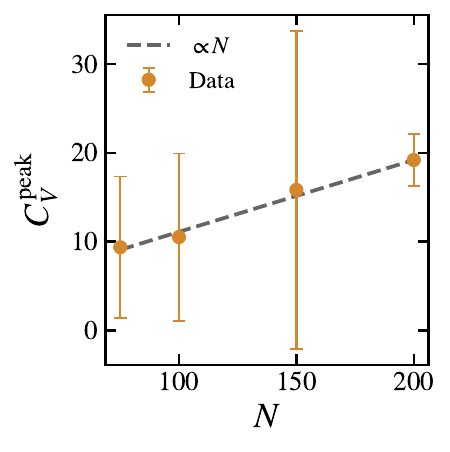}
        \caption{}
        \label{fig:gc_Cvmax}
    \end{subfigure}
    \hfill
    \begin{subfigure}[t]{0.32\textwidth}
        \centering
        \includegraphics[width=\linewidth]{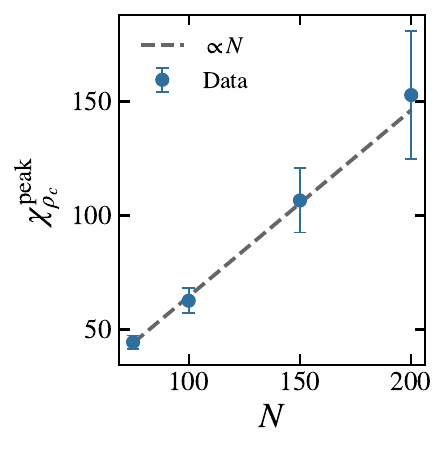}
        \caption{}
        \label{fig:gc_chirho_max}
    \end{subfigure}
    \hfill
    \begin{subfigure}[t]{0.32\textwidth}
        \centering
        \includegraphics[width=\linewidth]{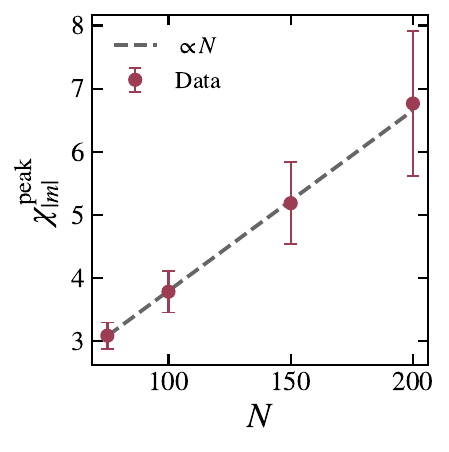}
        \caption{}
        \label{fig:gc_chim_max}
    \end{subfigure}

\caption{
\textbf{Size dependence of the maxima of the response functions associated with the Co--G transition.}
Maximum of the heat capacity $C_V^{\max}$ \textbf{(a)}, of the susceptibility associated with the contact density $\chi_{\rho_c}^{\max}$ \textbf{(b)}, and of the magnetic susceptibility $\chi_{|m|}^{\max}$ \textbf{(c)}, as a function of the chain length $N$, at $\mathcal K=6$, approaching from below. The circles represent the mean values obtained over the independent replicas, and the error bars their standard deviations. The dashed gray lines indicate the reference linear dependence $X_{\max}\propto N$.
}
    \label{fig:gc_peak_scaling}
\end{figure*}

We complement the distribution analysis with a finite-size study at $\mathcal K=6$. For each chain length, the observables and response functions are estimated from $50$ independent replicas. In the vicinity of the transition, the temperature is sampled with a step $\Delta\mathcal T=0.01$.
We examine the size dependence of the maxima of the heat capacity $C_V$, the susceptibility associated with the contact density $\chi_{\rho_c}$, and the magnetic susceptibility $\chi_{|m|}$. These three quantities respectively characterize the energetic, structural, and magnetic fluctuations accompanying the transition. The three maxima increase with chain length (Fig.~\ref{fig:gc_peak_scaling}). Their growth, close to a linear dependence on $N$, is consistent with the extensive behavior expected at a first-order transition. The agreement between several response functions thus provides a robust characterization of the finite-size dependence of the transition.

Additional finite-size diagnostics are presented in
Appendix~\ref{app:finite_size_diagnostics}. The sharpening of the
heat-capacity peak and the development of a pronounced minimum in the
Binder cumulant of the nematic order parameter with increasing $N$
provide complementary signatures consistent with first-order behavior.

\subsubsection{Globule--Crystal transition line}
The G--Cr transition separates the disordered globule from the crystalline phase and is located primarily from the maximum of the nematic susceptibility $\chi_S$.

\begin{figure*}[t]
    \centering

    \begin{subfigure}[t]{0.32\textwidth}
        \centering
        \includegraphics[width=\linewidth]{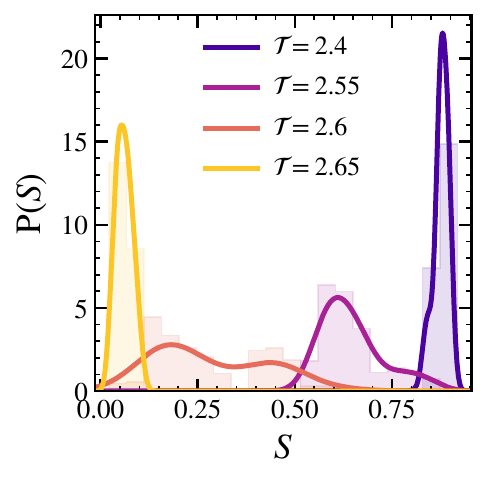}
        \label{fig:go_hist_S}
    \end{subfigure}
    \hfill
    \begin{subfigure}[t]{0.32\textwidth}
        \centering
        \includegraphics[width=\linewidth]{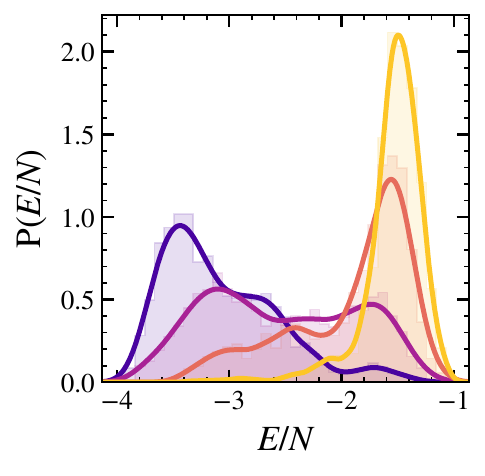}
        \label{fig:go_hist_energy}
    \end{subfigure}
    \hfill
    \begin{subfigure}[t]{0.32\textwidth}
        \centering
        \includegraphics[width=\linewidth]{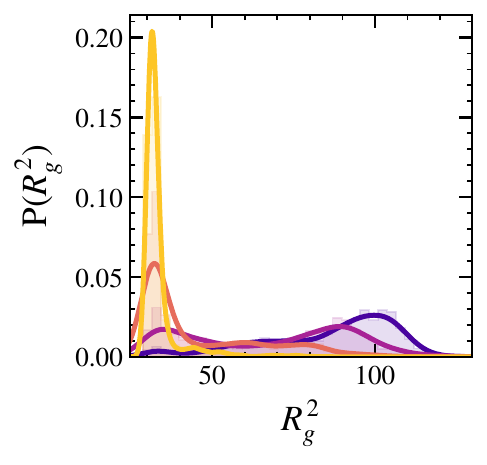}
        \label{fig:go_hist_Rg2}
    \end{subfigure}

    \caption{
\textbf{Distributions of the observables in the vicinity of the globule--crystal transition, for $N=600$ and $\mathcal K=6$.} Distributions of the nematic order parameter $S$ \textbf{(a)}, the energy per monomer $E/N$ \textbf{(b)}, and the squared radius of gyration $R_g^2$ \textbf{(c)}, approaching the critical value from below. The distributions correspond to $\mathcal T=2.4$ (purple), $2.55$ (magenta), $2.60$ (orange), and $2.65$ (yellow). The normalized histograms are complemented by Gaussian kernel density estimates (KDE).
    }
    \label{fig:go_distributions}
\end{figure*}

For $\mathcal K=6$ and $N=600$, the average of the pseudocritical temperatures estimated from the independent replicas is $\mathcal T_{\mathrm{GCr}}=2.75\pm0.05$. We examine the distributions of the nematic order parameter $S$, the energy per monomer $E/N$, and the squared radius of gyration $R_g^2$ when approaching the transition from below (Fig.~\ref{fig:go_distributions}).
At $\mathcal T=2.4$, the distribution $P(S)$ is concentrated at large values of $S$, characteristic of the crystalline phase. As the temperature increases, its maximum shifts toward lower values and the distribution broadens. This broadening becomes particularly pronounced at $\mathcal T=2.60$.
Over the same temperature interval, the distributions of $E/N$ and $R_g^2$ also broaden. These changes reflect a simultaneous increase in energetic and structural fluctuations, accompanied by a decrease in nematic order. The G--Cr transition is thus characterized by a collective reorganization of the polymer, with broadly distributed observables in the vicinity of the pseudocritical temperature.

\begin{figure}[t]
    \centering
    \includegraphics[width=0.8\linewidth]{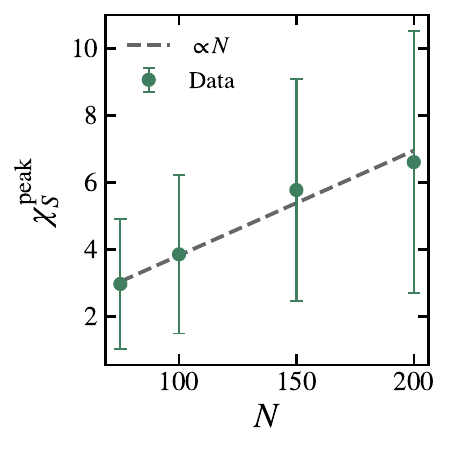}

    \caption{
    \textbf{Finite-size analysis of the nematic response.}
    Nematic susceptibility $\chi_S$ as a function of the reduced
    temperature $\mathcal T$ for different chain lengths $N$ at
    $\mathcal K=6$ \textbf{(a)}, and evolution of its maximum
    $\chi_S^{\max}$ with $N$ \textbf{(b)}.
    The dashed gray line indicates the
    reference linear dependence $\chi_S^{\max}\propto N$.
    }
    \label{fig:gcr_finite_size}
\end{figure}

We complement the distribution analysis with a finite-size study of the nematic susceptibility at $\mathcal K=6$. The maximum of $\chi_S$ becomes more pronounced as the chain length $N$ increases (Fig.~\ref{fig:gcr_finite_size}).

The growth of $\chi_S^{\mathrm{max}}$ with $N$, close to an extensive dependence over the sizes studied, is consistent with the behavior expected for a first-order transition.

\subsubsection{Coil--Crystal transition line}

At high stiffness, the maxima of $\chi_{\rho_c}$ and $\chi_S$ occur at the same temperature, indicating that compaction and the establishment of nematic order take place simultaneously.

For $\mathcal K=10$, the disordered globule is no longer resolved between the paramagnetic \textit{coil} and the crystalline phase. For $N=600$, the maxima of $\chi_S$ give $\mathcal T_c^{(S)}=3.45\pm0.05$.

\begin{figure}[t]
    \centering

    \begin{subfigure}[t]{0.48\columnwidth}
        \centering
        \includegraphics[width=\linewidth]{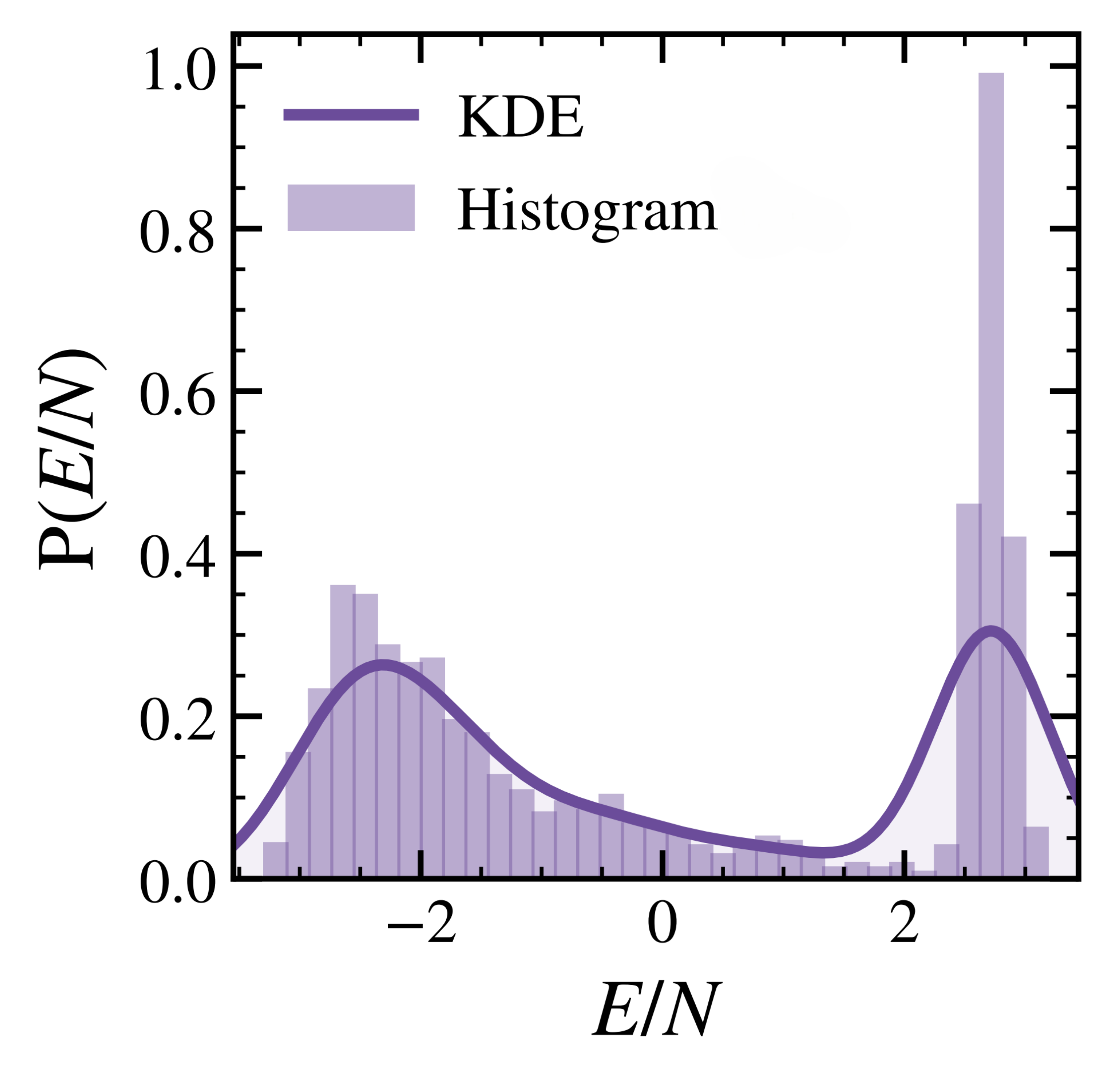}
        \caption{}
        \label{fig:cocr_k10_hist_energy}
    \end{subfigure}
    \hfill
    \begin{subfigure}[t]{0.48\columnwidth}
        \centering
        \includegraphics[width=\linewidth]{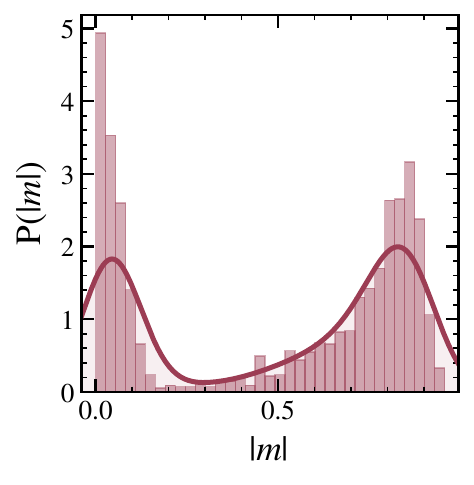}
        \caption{}
        \label{fig:cocr_k10_hist_m}
    \end{subfigure}

    \vspace{0.15cm}

    \begin{subfigure}[t]{0.48\columnwidth}
        \centering
        \includegraphics[width=\linewidth]{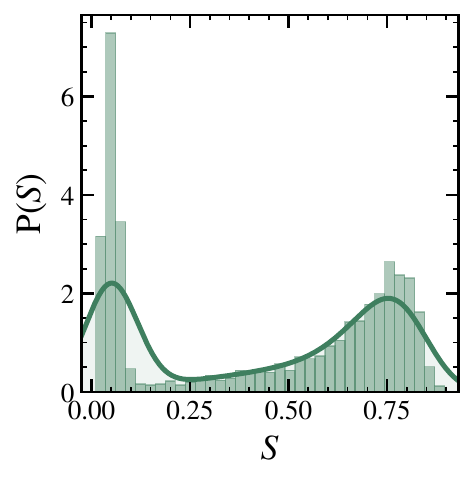}
        \caption{}
        \label{fig:cocr_k10_hist_S}
    \end{subfigure}
    \hfill
    \begin{subfigure}[t]{0.48\columnwidth}
        \centering
        \includegraphics[width=\linewidth]{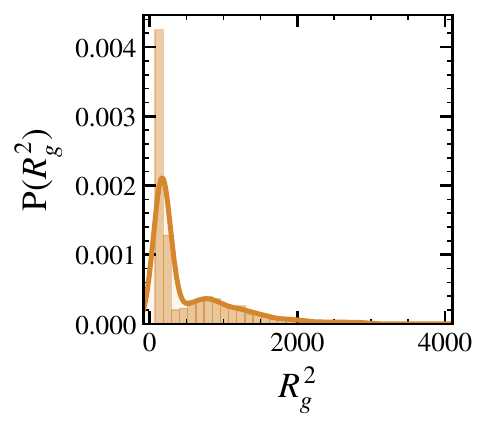}
        \caption{}
        \label{fig:cocr_k10_hist_Rg2}
    \end{subfigure}

    \caption{
    \textbf{Distributions of the observables near the direct
    coil--crystal transition.}
    Distributions of \textbf{(a)} the energy per monomer $E/N$,
    \textbf{(b)} the absolute magnetization $|m|$,
    \textbf{(c)} the nematic order parameter $S$, and
    \textbf{(d)} the squared radius of gyration $R_g^2$,
    for $N=600$, $\mathcal K=10$, and $\mathcal T=3.4$.
    Normalized histograms are complemented by Gaussian kernel density
    estimates (KDE).
    }
    \label{fig:cocr_k10_distributions}
\end{figure}

The agreement between these two estimates, within their uncertainties, identifies a direct Co--Cr transition combining chain compaction and the establishment of nematic order. For a chain of $N=600$ monomers, the distributions of the energy per monomer $E/N$, the absolute magnetization $|m|$, the nematic order parameter $S$, and the squared radius of gyration $R_g^2$ are examined at $\mathcal T=3.4$, when approaching the transition with increasing temperature (Fig.~\ref{fig:cocr_k10_distributions}).

The distributions reveal a simultaneous reorganization of the energetic, magnetic, and structural properties of the chain. The redistribution of $P(E/N)$ is accompanied by changes in $P(|m|)$ and $P(S)$, while $P(R_g^2)$ reflects the conformational change associated with the transition from the coil to the crystalline phase.

The coincidence of the two pseudocritical temperatures, together with the simultaneous redistribution of the energetic and structural observables and the bimodality observed for $N=200$, is consistent with a direct first-order Co--Cr transition.

\begin{figure}[t]
    \centering
    \begin{subfigure}[t]{0.48\linewidth}
        \centering
        \includegraphics[width=\linewidth]{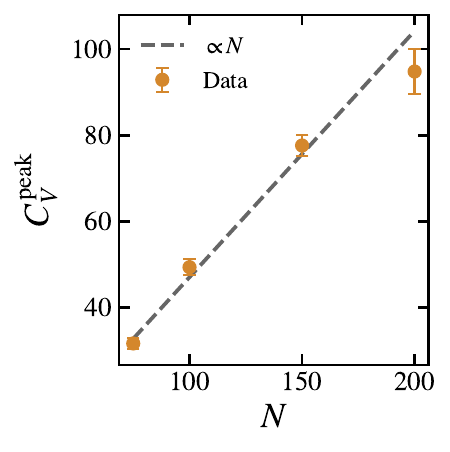}
        \caption{}
        \label{fig:cocr_k10_cv_scaling}
    \end{subfigure}
    \hfill
    \begin{subfigure}[t]{0.48\linewidth}
        \centering
        \includegraphics[width=\linewidth]{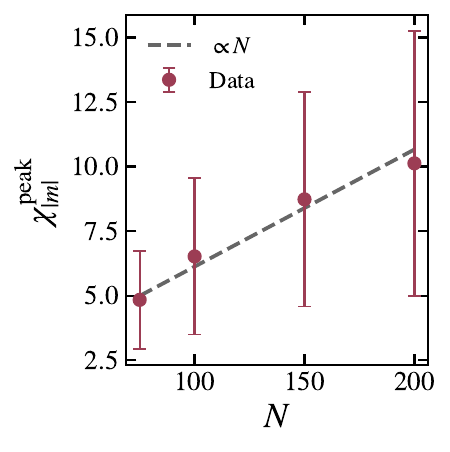}
        \caption{}
        \label{fig:cocr_k10_chim_scaling}
    \end{subfigure}

    \vspace{0.15cm}

    \begin{subfigure}[t]{0.48\linewidth}
        \centering
        \includegraphics[width=\linewidth]{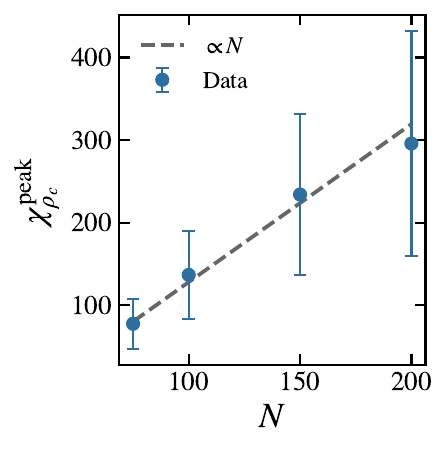}
        \caption{}
        \label{fig:cocr_k10_chirho_scaling}
    \end{subfigure}
    \hfill
    \begin{subfigure}[t]{0.48\linewidth}
        \centering
        \includegraphics[width=\linewidth]{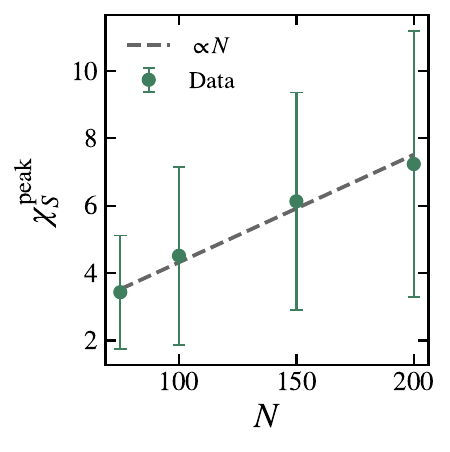}
        \caption{}
        \label{fig:cocr_k10_chiS_scaling}
    \end{subfigure}

    \caption{
    \textbf{Finite-size analysis of the direct coil--crystal transition
    at $\mathcal K=10$.}
    Evolution with chain length $N$ of the maxima of
    \textbf{(a)} the heat capacity $C_V^{\max}$,
    \textbf{(b)} the magnetic susceptibility $\chi_{|m|}^{\max}$,
    \textbf{(c)} the contact-density susceptibility
    $\chi_{\rho_c}^{\max}$, and
    \textbf{(d)} the nematic susceptibility $\chi_S^{\max}$.
    The maxima are obtained from response functions reconstructed from
    moments averaged over independent replicas. Dashed lines indicate
    the reference linear scaling $X_{\max}\propto N$ expected for a
    first-order transition.
    }
    \label{fig:cocr_k10_fss}
\end{figure}

We complement the distribution analysis with a finite-size study of the maxima of the response functions associated with the direct Co--Cr transition.
Figure~\ref{fig:cocr_k10_fss} shows the evolution with chain length of the maxima of the heat capacity \(C_V\), the magnetic susceptibility \(\chi_{|m|}\), the susceptibility associated with the contact density \(\chi_{\rho_c}\), and the nematic susceptibility \(\chi_S\).
For all four observables, the peak height increases with \(N\). This evolution reflects the strengthening of energetic, magnetic, and structural fluctuations as the system size increases. Over the accessible range of sizes, the data are consistent with an approximately linear growth with \(N\), as expected at leading order for a first-order transition in a finite system.

The simultaneous growth of the four maxima shows that the transition couples the energetic, magnetic, and structural degrees of freedom. It complements the coincidence of the pseudocritical temperatures and the signatures observed in the distributions.
Additional finite-size evidence based on the Binder cumulant is presented in Appendix~\ref{app:finite_size_diagnostics}.
Taken together, these observations are consistent with a direct first-order Co--Cr transition.

\section{Analytical results}

\subsection{Free-energy upper-bound construction}

To rationalize the zero-field phase diagram obtained from the simulations, we construct approximate upper bounds for the free energy of the three competing phases: the swollen coil, the disordered globule, and the ordered crystalline phase. Building on the free-energy upper-bound strategy used for the magnetic polymer~\cite{Garel1999PhasePolymers}, we extend the coil and globule constructions by incorporating the FCC bending contribution, while the crystalline branch is introduced separately through a geometrically motivated ansatz. A full Landau-type treatment of the semiflexible magnetic polymer would require several coupled fields because density, magnetization, local curvature, and orientational order are coupled. We therefore use compact phase-wise upper bounds that retain the leading competition between conformational entropy, spin-mediated attraction, and bending stiffness.

We first construct, for each phase, a restricted and factorized partition function containing the dominant contributions expected for that class of conformations. Since restricting the available configurations gives a lower estimate of the partition function, the corresponding quantity $F_X=-T\ln Z_X$ provides an upper bound for the free energy for the $X$ phase. The phase-wise bounds can then be compared on the same footing. 

We use the reduced free-energy density $f_X=F_X/(JN)$, with $X\in\{\mathrm{coil},\mathrm{globule},\mathrm{crystal}\}$. We also set $k_{\mathrm B}=1$, so that $\beta=1/T$.

\subsection{Free-energy bounds for the three phases}

Within this construction, the coil is represented as an FCC self-avoiding walk, with its magnetic contribution approximated by a one-dimensional Ising chain, while the globule is represented by a Hamiltonian path on the FCC lattice with its corresponding Hamiltonian path entropy estimate $(q/e)^N$~\cite{Orland1985Hamiltonian}. 

Since the reversal of the preceding bond is forbidden, on the FCC lattice there are eleven possible local continuations: one straight continuation, four bends of $60^\circ$, two of $90^\circ$, and four of $120^\circ$. Their normalized Boltzmann weighting gives
\begin{equation}
    w_{\mathrm{FCC}}(\mathcal{K},\mathcal{T})
    =
    \frac{
    1+4e^{-\mathcal{K}/(2\mathcal{T})}
    +2e^{-\mathcal{K}/\mathcal{T}}
    +4e^{-3\mathcal{K}/(2\mathcal{T})}
    }{11},
    \label{eq:analytical_bending_factor}
\end{equation}
which approaches unity in the flexible limit ($\mathcal{K} \to 0$). This factor is a local Boltzmann reweighting of the conformations already counted by the underlying polymer ensemble.

For the coil, the number of FCC self-avoiding conformations scales as $\mu^N N^{\gamma-1}$, where $\mu\simeq10.037$ is the FCC connective constant and $\gamma$ is the self-avoiding-walk entropic exponent~\cite{Guttmann1989Critical,Schram2017Enumeration}. In an extended conformation, non-consecutive spatial contacts are rare, and the magnetic contribution is approximated by the one-dimensional Ising factor $2\left[2\cosh(\beta J)\right]^{N-1}$. Including the bending reweighting at the $N-2$ internal joints therefore gives the trial partition function
\begin{equation}
    Z_{\mathrm{coil}}
    \sim
    \mu^N N^{\gamma-1}
    \left[2\cosh(\beta J)\right]^{N-1}
    w_{\mathrm{FCC}}^{\,N-2}.
    \label{eq:analytical_zcoil}
\end{equation}
The algebraic factor $N^{\gamma-1}$, overall multiplicative constants, and the distinction between $N$, $N-1$ and $N-2$ are subleading in the thermodynamic limit.

For the globule, the polymer is instead represented by a compact Hamiltonian path. The number of such compact conformations is approximated by $\left(q/e\right)^N$, with $q=12$ for the FCC lattice. In a fully magnetized compact bulk, each monomer has coordination number $q=12$, and because each contact is shared by two monomers, division by two gives an interaction energy $-qJ/2=-6J$ per monomer. Finite compact structures receive positive surface corrections because surface monomers have fewer than twelve neighbours, but these corrections vanish per monomer in the thermodynamic limit. Using, as a first approximation, the same local FCC bending factor for the globule as for the coil, the thermodynamic contribution of the globule is therefore
\begin{equation}
    Z_{\mathrm{globule}}
    \sim
    \left(\frac{q}{e}\right)^N
    e^{6\beta JN}
    w_{\mathrm{FCC}}^{\,N-2}.
    \label{eq:analytical_zglobule}
\end{equation}
The use of the same local bending factor for the coil and globule, together with its comparison with measured bend statistics, is discussed further in Appendix~\ref{app:analytical_bounds_1}.

The crystalline phase is also compact. The zero-temperature analysis in Sec.~\ref{subsec:low_energy_stiff} showed how the geometry of dense folded structures reorganizes through the competition between favorable contacts and bending cost, while the corresponding surface and bending corrections remain subextensive. Here, we use this observation only to construct a variational upper bound for the crystalline free energy in the thermodynamic limit. We consider an admissible maximally compact ordered FCC conformation at zero temperature, without specifying its detailed morphology. It may, for example, be realized as a compact folded or toroidal structure; the particular construction is immaterial for the leading bulk contribution.

In the compact FCC bulk, each monomer has coordination number $q=12$.
Since each pair interaction is shared by two monomers, a fully
magnetized bulk contributes
\begin{equation}
E_0^{\mathrm{bulk}}=-\frac{q}{2}JN=-6JN.
\end{equation}
For a finite compact structure, monomers at the boundary have fewer
neighbours, while the polymer backbone must also bend in order to remain
confined. For an ordered compact sequence in which these defects are
restricted to the surface or turning regions, both contributions are
subextensive. We may therefore write
\begin{equation}
E_0^{\mathrm{crystal}}
=
-6JN+\Delta E_0^{\mathrm{surf}}+\kappa n_b,
\end{equation}
with
\begin{equation}
\frac{\Delta E_0^{\mathrm{surf}}}{N}\rightarrow0,
\qquad
\frac{n_b}{N}\rightarrow0
\qquad (N\rightarrow\infty).
\end{equation}
Consequently,
\begin{equation}
f_{\mathrm{crystal}}
=
\lim_{N\to\infty}\frac{F_{\mathrm{crystal}}}{JN}
=
-6.
\end{equation}
This construction does not assert a unique microscopic morphology for
the crystalline phase. It only provides one admissible compact ordered
family and hence a free-energy upper bound; any additional crystalline
configurations or finite-temperature defects can only enlarge the
restricted partition function and lower the corresponding free energy.

Taking $-T\ln Z_X/(JN)$ and retaining the leading thermodynamic-limit contributions gives the following free energy bounds

\begin{equation}
\begin{alignedat}{2}
&f_{\mathrm{coil}} \quad
&&= -\mathcal{T}\ln\mu
-\mathcal{T}\ln\!\left[2\cosh\!\left(\frac{1}{\mathcal{T}}\right)\right]
-\mathcal{T}\ln w_{\mathrm{FCC}},
\\[0.5ex]
&f_{\mathrm{globule}} \quad
&&= -6
-\mathcal{T}\ln\!\left(\frac{q}{e}\right)
-\mathcal{T}\ln w_{\mathrm{FCC}},
\\[0.5ex]
&f_{\mathrm{crystal}} \quad
&&= -6.
\end{alignedat}
\label{eq:analytical_free_energy_bounds}
\end{equation}

\begin{figure}[t]
    \centering
    \includegraphics[width=0.9\columnwidth]{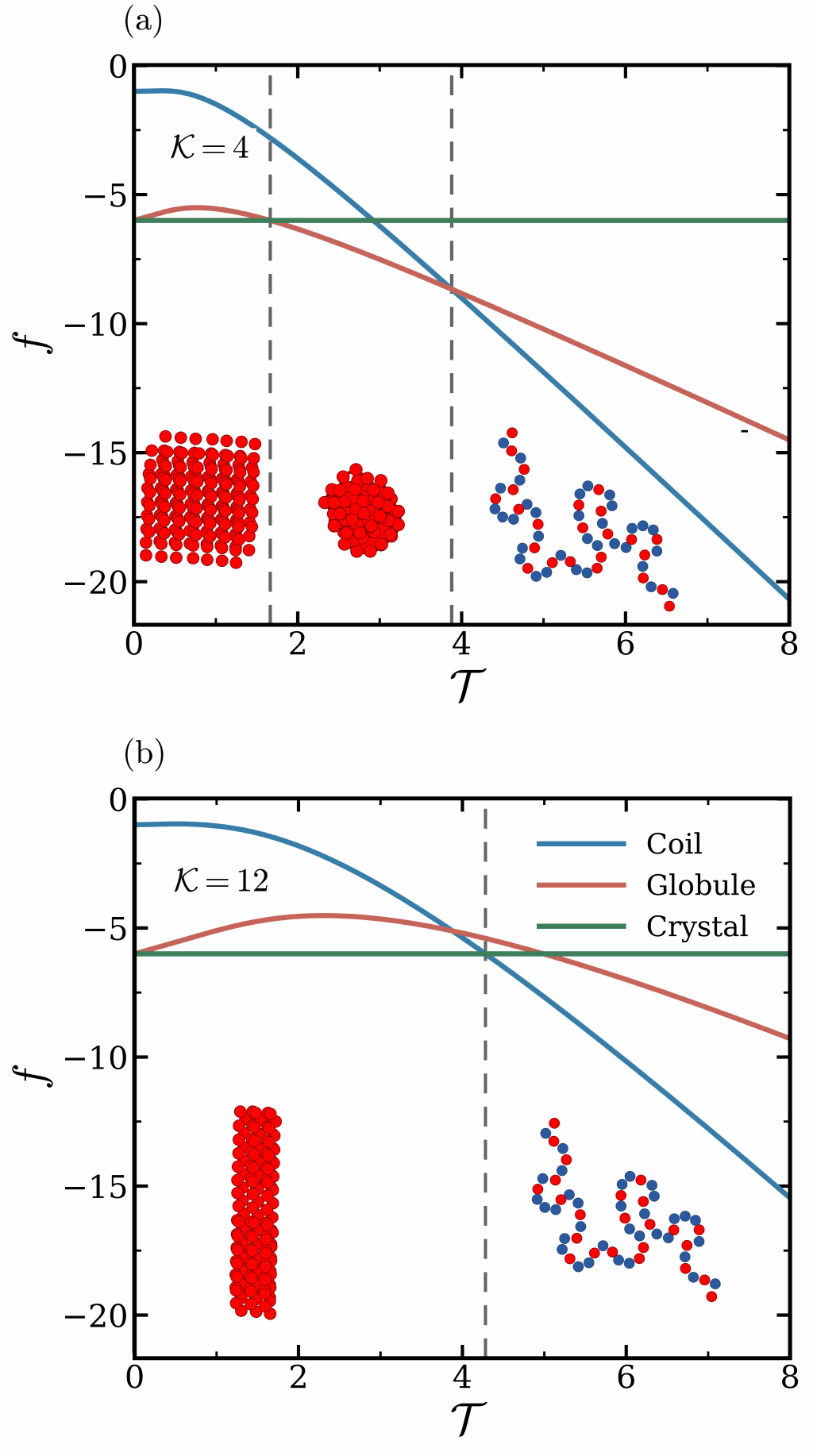}
    \caption{\textbf{Thermodynamic-limit free-energy upper bounds $f=F/(JN)$ for the coil, globule, and crystalline phases as functions of temperature, shown for two representative bending rigidities.} For $\mathcal K=4$ \textbf{(a)}, the lowest bound changes from crystal at low temperature to globule at intermediate temperature and finally to coil at high temperature. RFor $\mathcal K=12$ \textbf{(b)}, the globule no longer becomes the lowest free-energy branch, yielding a direct crystal--coil transition.}
    \label{fig:analytical_bounds}
\end{figure}

\subsection{Thermodynamic limit phase diagram}

The competition between the three bounds is illustrated in Fig.~\ref{fig:analytical_bounds}. At low to moderate stiffness, the three phases appear successively as temperature is increased: the crystal bound is lowest at low temperature, followed by the globule and finally the coil. At sufficiently large stiffness, the bending cost raises the globule branch relative to the crystal, and the intermediate globular regime disappears. The system then passes directly from the crystalline to the coil branch.

\begin{figure}[t]
    \centering
    \includegraphics[width=0.9\columnwidth]{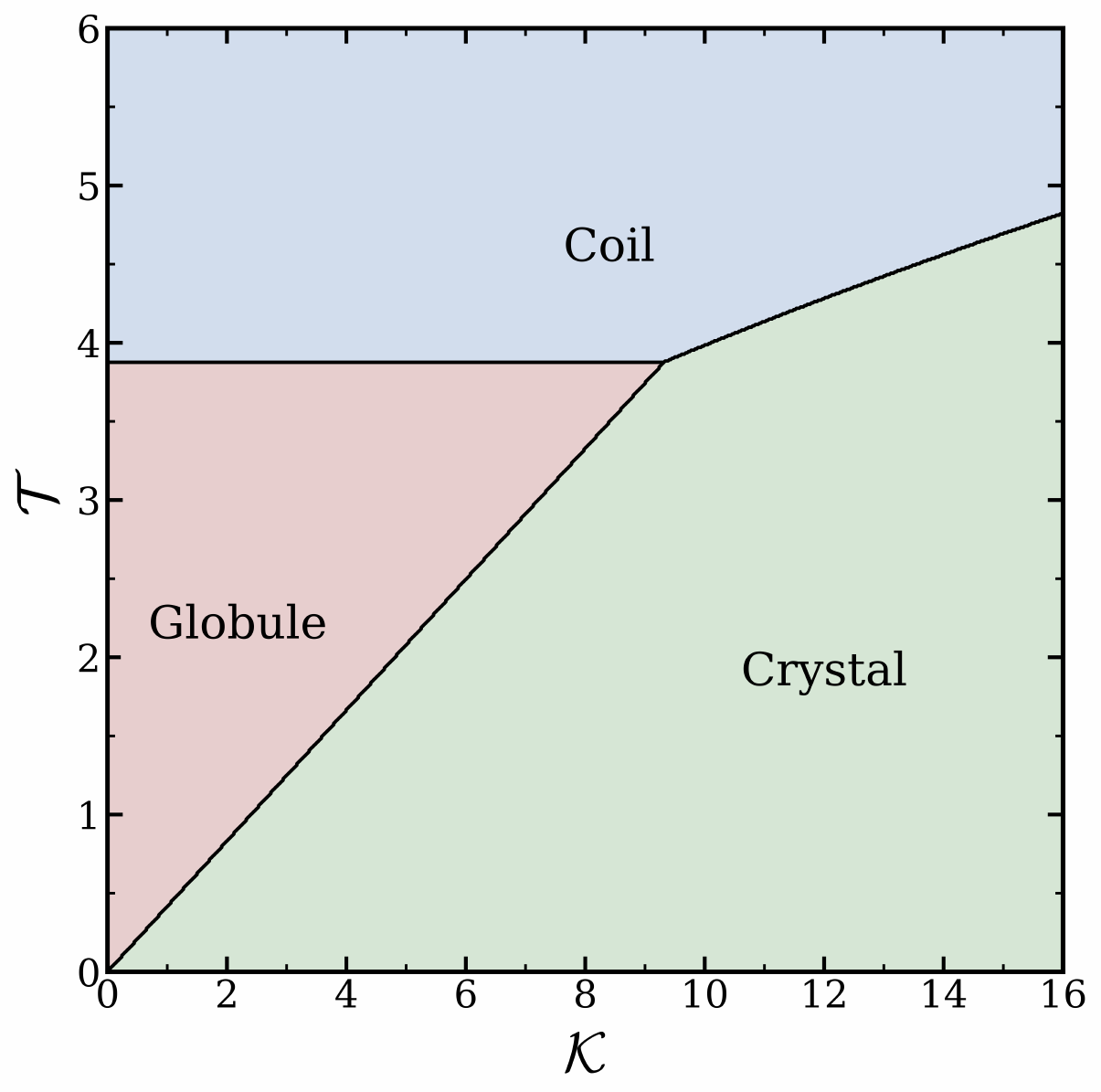}
    \caption{\textbf{Thermodynamic-limit analytical phase diagram at zero field}. The phase diagram was obtained by selecting the lowest of the three free-energy bounds in Eq.~\eqref{eq:analytical_free_energy_bounds}. The theory predicts coil, globule, and crystal regions, with the globule eliminated beyond the triple point in favour of a direct crystal--coil transition.}
    \label{fig:analytical_phase_diagram}
\end{figure}

The analytical phase diagram is obtained by pairwise matching the bounds and selecting the smallest one at each $(\mathcal K,\mathcal T)$, as shown in Fig.~\ref{fig:analytical_phase_diagram}. Because the same local bending factor enters the coil and globule bounds, it cancels from their equality, producing a horizontal coil--globule boundary in this thermodynamic-limit approximation. The triple point satisfies
\begin{align}
\ln\mu+\ln\!\left[2\cosh\!\left(\frac{1}{\mathcal{T}_\ast}\right)\right]
    &=
    \frac{6}{\mathcal{T}_\ast}
    +\ln\!\left(\frac{q}{e}\right),
    \nonumber\\
w_{\mathrm{FCC}}(\mathcal{K}_\ast,\mathcal{T}_\ast)
    &=\frac{e}{q},
    \label{eq:analytical_triple_point}
\end{align}
which gives
\begin{equation}
\mathcal{T}_\ast\simeq3.9,
\qquad
\mathcal{K}_\ast\simeq9.3.
\end{equation}

The upper-bound construction reproduces the overall three-phase topology of the Monte Carlo phase diagram: coil, globule, and crystal phases meeting at a triple region, followed by a direct crystal--coil boundary at sufficiently large stiffness. Some quantitative differences remain, most notably the slight stiffness dependence of the finite-chain coil--globule boundary. These may arise from compact-confinement and finite-size corrections to the globule's local bend statistics that are absent in the present thermodynamic-limit approximation.

\section{Discussion and conclusions}

We have investigated the phase diagram of a semiflexible magnetic polymer, highlighting the interplay between magnetic degrees of freedom, attractive interactions, and chain stiffness. In the era of high-resolution, high-throughput techniques, there is an increasing need for models of nucleoprotein complexes at the scale of the DNA persistence length {\it in vivo}, in order to understand how bridging proteins accommodate the intrinsic stiffness of DNA.

The morphology of the crystalline phase (ground state) can be understood as the result of a competition between contact energy and bending cost. In compact ferromagnetic states, spin alignment makes the interaction between contacting monomers essentially homogeneous, so that the selection of structures involves mechanisms similar to those known for semiflexible homopolymers. Increasing stiffness favors longer straight segments and leads to more anisotropic and elongated structures. The nematic order observed in the crystalline phase is thus directly related to this geometry. This evolution also shows that the radius of gyration alone is not a sufficient measure of compaction: a structure can remain highly compact while exhibiting a large $R_g$ as it becomes anisotropic.

We find that increasing stiffness progressively stabilizes the crystalline phase at the expense of the intermediate globular phase, eventually leading to a direct transition between the coil and crystalline phases. The magnetic degrees of freedom introduce additional ordering phenomena within this overall conformational organization, with ferromagnetic order emerging upon compaction and nematic order developing in the crystalline regime. Stiffness therefore primarily affects the relative stability of the globular and crystalline phases, while the coil--globule boundary remains comparatively insensitive to stiffness.

This overall organization of the phase diagram is reminiscent of that of an attractive semiflexible homopolymer. In the latter case, increasing stiffness also progressively stabilizes crystalline structures at the expense of the globular phase. Despite the presence of magnetic degrees of freedom, the overall sequence of regimes with increasing stiffness is therefore qualitatively preserved. The analogy, however, does not extend trivially to the nature of the transitions. In the semiflexible magnetic polymer, the coil--globule transition is associated with the simultaneous onset of compaction and ferromagnetic order, leading to a first-order transition, whereas the semiflexible homopolymer displays a coil-to-globule transition corresponding to a second-order $\theta$-like transition. The magnetic degrees of freedom thus provide an additional ordering mechanism that is coupled to the conformational transitions of the chain.

The magnetic degrees of freedom therefore modify the nature of the transitions. The coil--globule transition remains associated with the simultaneous onset of compaction and ferromagnetic order, with signatures compatible with a discontinuous transition, as in the flexible magnetic polymer. Within the range investigated, stiffness does not appear to decouple compaction from the emergence of magnetic order. The globule--crystal transition is accompanied by a reorganization of nematic order and is likewise compatible with a discontinuous scenario. Highly degenerate, long-lived metastable states are observed in the simulations. However, the finite-size analysis of the various susceptibilities and of the heat capacity provides unambiguous evidence for first-order transitions. Likewise, the direct coil--crystal transition observed at high stiffness displays finite-size coexistence signatures, and its discontinuous character is confirmed by a systematic finite-size analysis.

The analytical description provides a compact thermodynamic interpretation of these results. Despite its approximations, it reproduces the overall topology of the phase diagram and shows that it can be understood in terms of the competition between conformational entropy, contact energy, and bending cost. However, the approximations introduced in the different branches, in particular the restricted description of the crystalline phase, for which configurational entropy and thermal defects are neglected at leading order, limit its quantitative predictive power. Despite these approximations, the analytical construction provides a thermodynamic description of the numerical phase diagram. It predicts a triple point at $(\mathcal K_\ast,\mathcal T_\ast)\simeq(9.3,3.9)$, close to the numerical triple point $(8,3.2)$, and gives transition temperatures in reasonable agreement with the Monte Carlo results. The construction should therefore be regarded as a description of the dominant stabilization mechanisms with useful semi-quantitative predictive power, rather than as an exact theory of the transition temperatures or orders. Natural improvements would include accounting for bending correlations in compact conformations and for thermal defects in the crystalline phase.

Several perspectives emerge from this work. A first concerns the limit of vanishing stiffness, for which the connection between the crystalline phase at finite stiffness and the strictly flexible limit remains an open question~\cite{Arcangeli2024PhaseBehaviorSemiflexiblePoorSolvent}. In our simulations, the globule--crystal boundary can no longer be directly resolved for a renormalized stiffness below two. Methods allowing a more efficient exploration of the different conformational basins, such as parallel tempering or Wang--Landau approaches, would make it possible to investigate this limit more systematically.

A second, particularly natural, perspective is to introduce a nonzero magnetic field. In the $h\rightarrow0$ limit, the magnetic degrees of freedom remain fluctuating, whereas in the strong-field limit the spins become polarized and the model approaches that of a semiflexible homopolymer. Studying the combined effects of field and stiffness would therefore make it possible to determine a possible critical line $h_c(\mathcal K)$ and construct the three-dimensional phase diagram $(\mathcal T,\mathcal K,h)$, connecting the fluctuating magnetic regime to the crystalline regimes stabilized by stiffness. Such an extension would in particular help clarify how the coupling between magnetic order, compaction, and stiffness modifies the nature of the transitions and the overall structure of the phase diagram.

\section{Appendices}

\subsection{Numerical methods}
\label{app:simulation_protocol}

For each value of $\mathcal K$, the heating simulations are initialized
from a low-energy configuration obtained at $\mathcal T=0$.
The temperature is then increased stepwise, using the final
configuration obtained at one temperature as the initial configuration
at the next temperature.

For each parameter set, simulations consist of a thermalization stage of
\[
t_{\rm therm}=9\times10^{4}\ \mathrm{MCS},
\]
followed by a production stage of
\[
t_{\rm sim}=3\times10^{6}\ {\rm MCS}.
\]
Observables are recorded every
\[
\tau=7.5\times10^{4}\ \mathrm{MCS}.
\]

For the finite-size analyses, each chain length is simulated using
$50$ independent replicas. In the vicinity of the transitions, the
temperature is sampled with a spacing $\Delta\mathcal T=0.01$.
Observables and response functions are obtained by averaging over the
independent replicas.

\subsection{Additional finite-size diagnostics}
\label{app:finite_size_diagnostics}

To complement the finite-size analyses presented in the main text, we
consider the temperature dependence of the heat capacity and the Binder
cumulant of the nematic order parameter.

\begin{figure}[!h]
    \centering
    \includegraphics[width=\columnwidth]{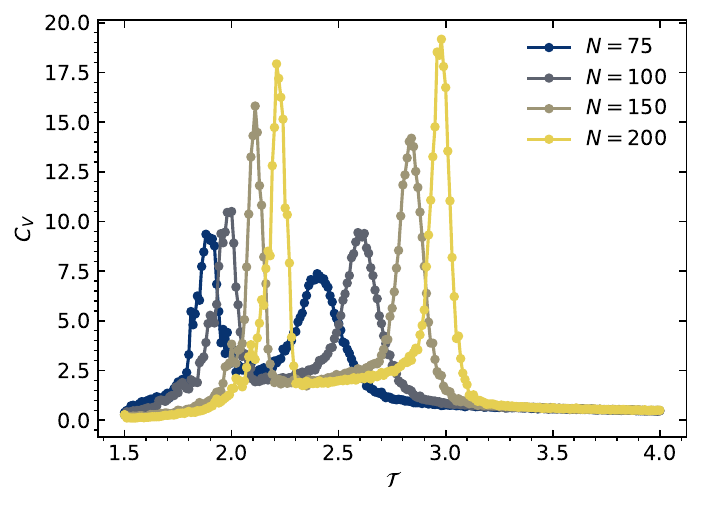}

    \caption{
    \textbf{Finite-size evolution of the heat capacity at $\mathcal K=6$.}
    Heat capacity $C_V$ as a function of the reduced temperature
    $\mathcal T$ for different chain lengths $N$.
    }
    \label{fig:Cv_k6}
\end{figure}

Figure~\ref{fig:Cv_k6} shows the heat capacity $C_V$ in the vicinity of
the G--Cr transition at $\mathcal K=6$ for different chain lengths $N$.
The peak becomes increasingly pronounced as $N$ increases, consistently
with the finite-size evolution of the response functions discussed in
the main text.

As an additional finite-size diagnostic, we consider the fourth-order
Binder cumulant of the nematic order parameter,
\begin{equation}
U_S
=
1-
\frac{\langle S^4\rangle}
     {3\langle S^2\rangle^2}.
\label{eq:binder_S}
\end{equation}

\begin{figure}[t]
    \centering

    \begin{subfigure}[t]{\columnwidth}
        \centering
        \includegraphics[width=0.85\linewidth]{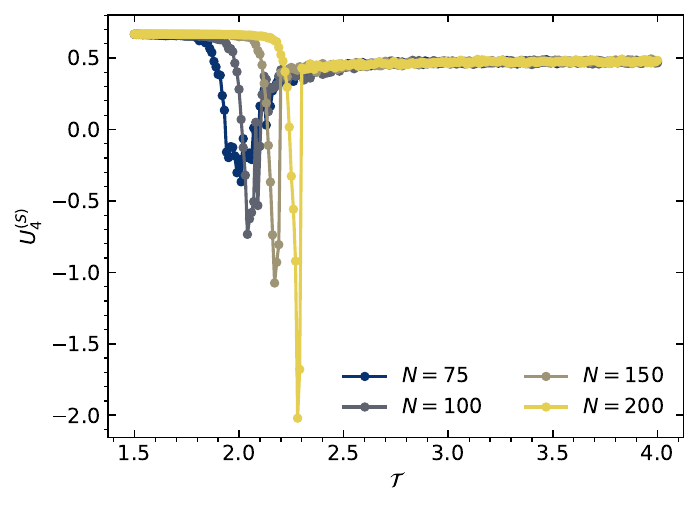}
        \caption{}
        \label{fig:binder_k6}
    \end{subfigure}

    \vspace{0.15cm}

    \begin{subfigure}[t]{\columnwidth}
        \centering
        \includegraphics[width=0.85\linewidth]{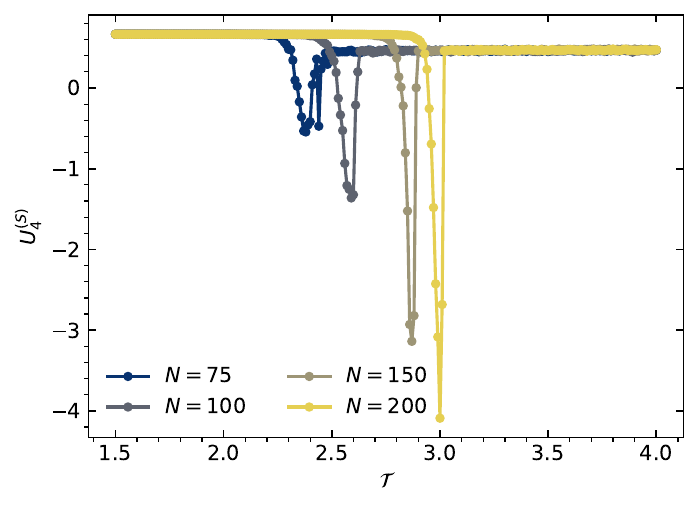}
        \caption{}
        \label{fig:binder_k10}
    \end{subfigure}

    \caption{
    \textbf{Finite-size evolution of the Binder cumulant of the nematic
    order parameter.}
    Binder cumulant $U_S$ as a function of the reduced temperature
    $\mathcal T$ for different chain lengths $N$.
    \textbf{(a)} Globule--crystal transition at $\mathcal K=6$.
    \textbf{(b)} Direct coil--crystal transition at $\mathcal K=10$.
    }
    \label{fig:binder_transitions}
\end{figure}

Figure~\ref{fig:binder_transitions} shows $U_S$ for the G--Cr transition
at $\mathcal K=6$ and for the direct Co--Cr transition at
$\mathcal K=10$. In both cases, the evolution of the Binder-cumulant
minimum with increasing chain length provides a complementary
finite-size diagnostic of the transition behavior.

\subsection{Validation of the FCC bending factor}
\label{app:analytical_bounds_1}

\begin{figure}[t]
    \centering
    \includegraphics[width=\columnwidth]{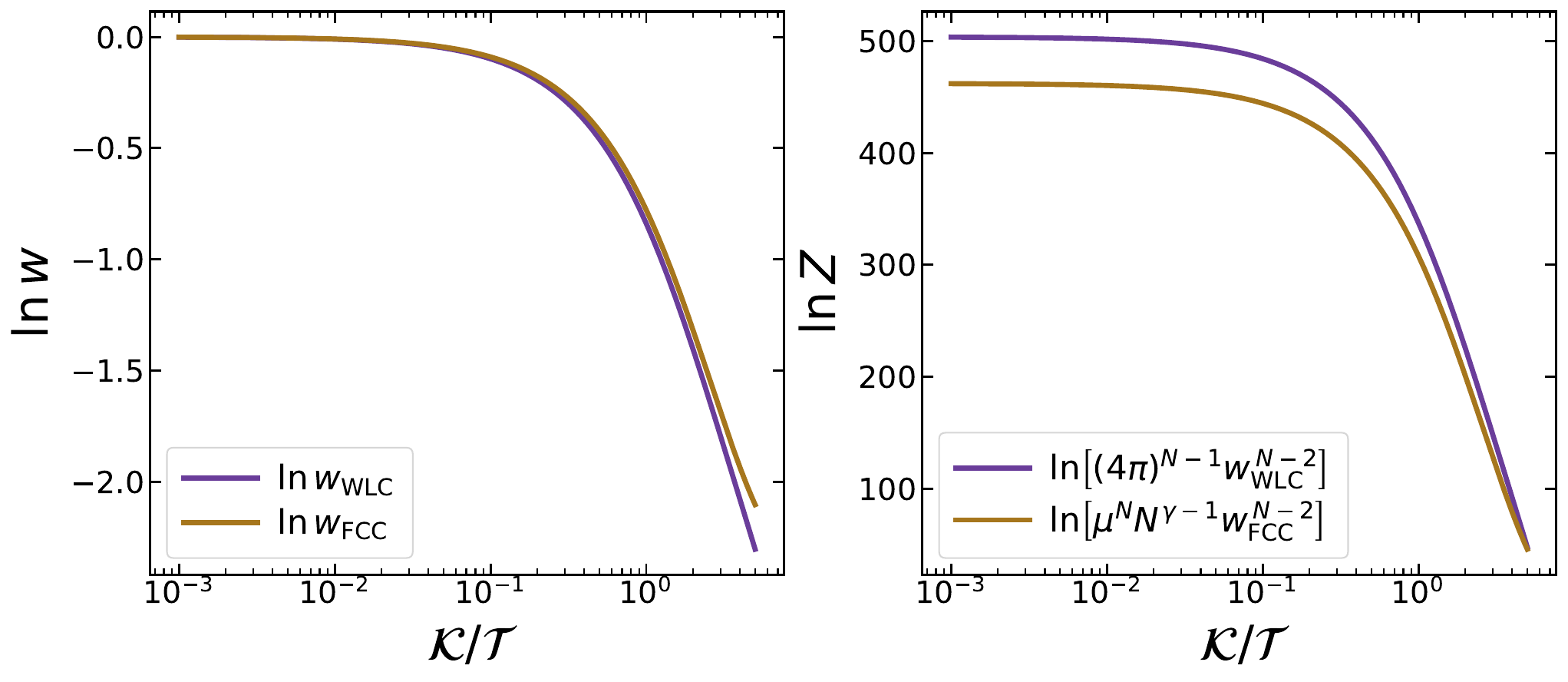}
    \caption{\textbf{Comparison of the FCC bending factor with the continuum WLC result.} The local factors and the corresponding full conformational measures display the same suppression with increasing bending stiffness.}
    \label{fig:annex_wlc_benchmark}
\end{figure}

The local FCC bending factor used in the coil and globule bounds can first be compared with the corresponding continuum worm-like-chain (WLC) result. With the same bending energy convention, the normalized single-joint WLC factor is
\begin{equation}
    w_{\mathrm{WLC}}(\mathcal{K},\mathcal{T})
    =
    \frac{1-e^{-2\mathcal{K}/\mathcal{T}}}
    {2\mathcal{K}/\mathcal{T}}.
    \label{eq:annex_wlc_factor}
\end{equation}
Both $w_{\mathrm{WLC}}$ and the FCC factor $w_{\mathrm{FCC}}$ defined in Eq.~\eqref{eq:analytical_bending_factor} approach unity in the flexible limit and decrease monotonically as bending becomes more costly~\cite{Marantan2018WormLike}. The comparison in Fig.~\ref{fig:annex_wlc_benchmark} shows that the explicit FCC enumeration reproduces the same qualitative suppression of bent configurations as the continuum angular integral. The same qualitative trend is obtained after combining the local factors with their respective conformational measures, $(4\pi)^{N-1}w_{\mathrm{WLC}}^{N-2}$ for the continuum chain and $\mu^N N^{\gamma-1}w_{\mathrm{FCC}}^{N-2}$ for the FCC self-avoiding walk~\cite{Marantan2018WormLike,Fiasconaro2023Elastic}.

\begin{figure}[t]
    \centering
    \includegraphics[width=\columnwidth]{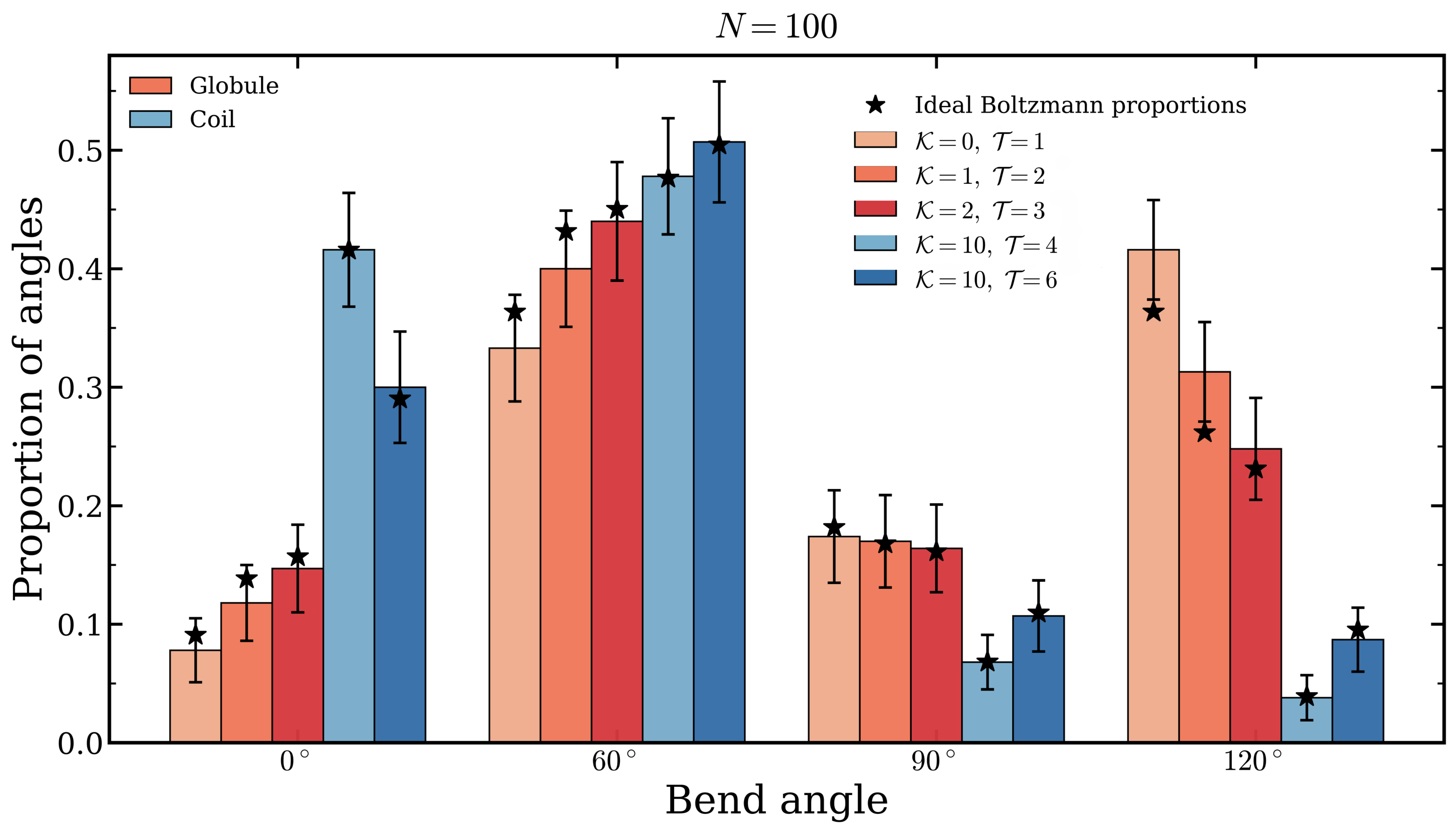}
    \caption{\textbf{Measured FCC bend-angle proportions for $N=100$ compared with the ideal local Boltzmann weights.} Both the coil and globule data agree well with the theoretical values.}
    \label{fig:annex_bend_angles}
\end{figure}

For the coil, the local FCC Boltzmann weights associated with bends of $0^\circ$, $60^\circ$, $90^\circ$, and $120^\circ$ are proportional to
\begin{equation}
    1,\qquad
    4e^{-\mathcal{K}/(2\mathcal{T})},\qquad
    2e^{-\mathcal{K}/\mathcal{T}},\qquad
    4e^{-3\mathcal{K}/(2\mathcal{T})},
    \label{eq:annex_bend_weights}
\end{equation}
respectively. Figure~\ref{fig:annex_bend_angles} compares these proportions with Monte Carlo measurements for chains of length $N=100$. Both the coil-regime and globule-regime data agree reasonably well with the ideal Boltzmann proportions for all four angles.

This agreement supports the use of the same local FCC bending factor for the coil and globule in the thermodynamic-limit free-energy construction. At the level of approximation considered here, the local bend angles in both equilibrium regimes are therefore well described by their FCC Boltzmann weights, while the two phases remain distinguished by their conformational entropies and magnetic interaction energies. Possible finite-size differences between the bend-angle statistics of swollen and compact FCC conformations, in particular for compact Hamiltonian paths, remain an interesting question for further study.

\subsection{Explicit construction of the crystalline upper bound}

The crystalline upper bound used to construct the thermodynamic limit phase diagram does not assume a
unique microscopic morphology. It only requires the existence of an
ordered, maximally compact sequence of FCC conformations for which the
loss of bulk contacts and the number of bends are subextensive in
\(N\). As an explicit realization of this compact sequence, we consider an ordered, compact cylindrical bundle filled by parallel strands of characteristic
length \(\ell\), connected by folds at the two end faces. The
cross-sectional area therefore scales as
\[
M\sim \frac{N}{\ell}.
\]

\begin{figure}[!h]
    \centering
    \includegraphics[width=0.75\columnwidth]{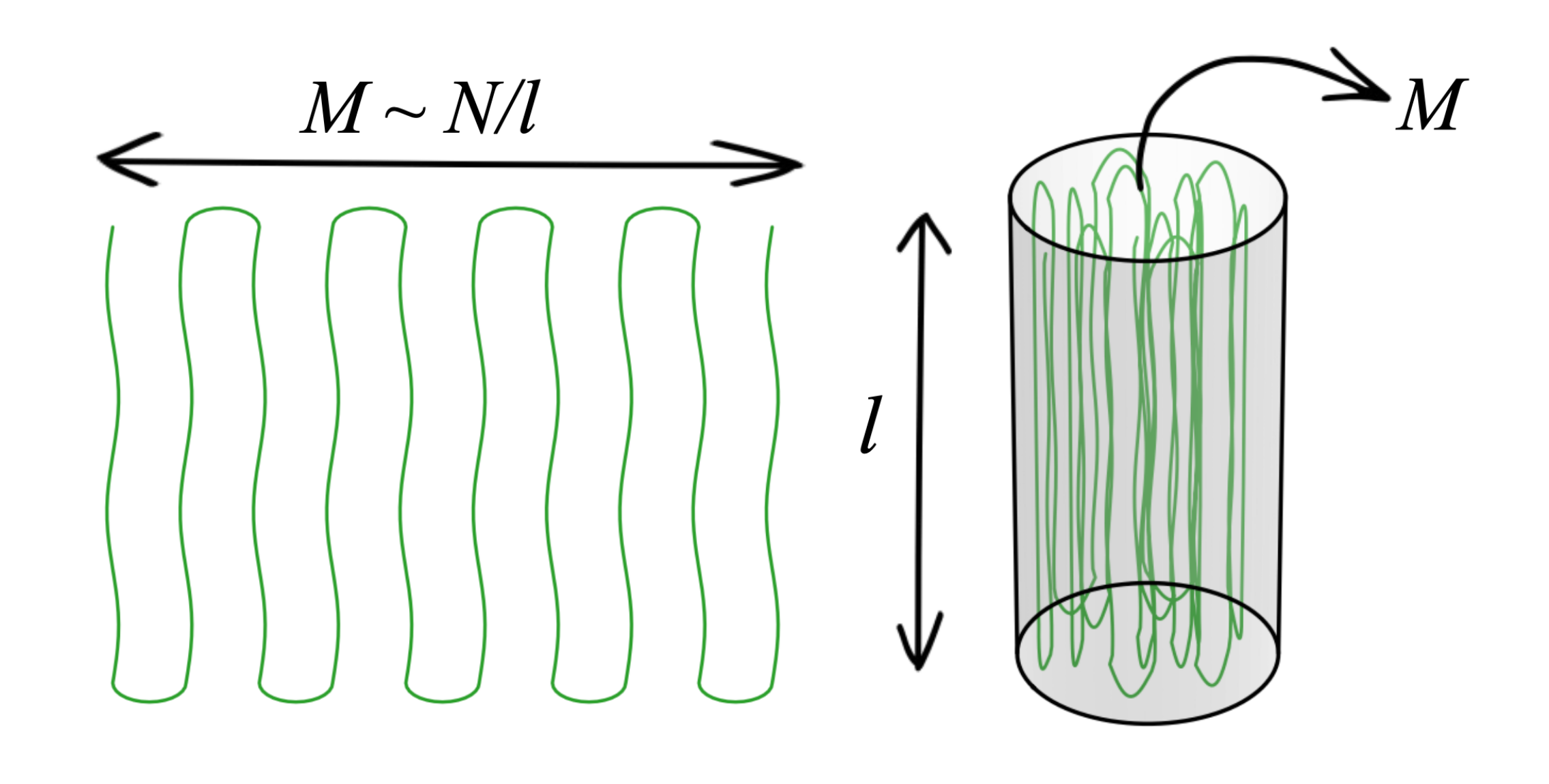}
    \caption{\textbf{Cynlindrical bundle construction for the crystalline phase.} The chain forms nearly parallel strands of length $\ell$, while bends are localized at the folded end faces of cross-sectional area $M\sim N/\ell$.}
    \label{fig:annex_folded_bundle}
\end{figure}

The leading bulk energy of a maximally compact FCC structure is
\(-6JN\). For the finite system, missing contacts on the
lateral surface scale as the lateral area,
\[
\Delta n_c^{\mathrm{lat}}\sim (N\ell)^{1/2},
\]
whereas missing contacts at the two end faces scale as
\[
\Delta n_c^{\mathrm{end}}\sim \frac{N}{\ell}.
\]
The folds are likewise localized at the end faces, giving
\[
n_b
=
2\left(\frac{N}{\ell}-1\right)
\sim \frac{2N}{\ell}.
\]
The zero-temperature finite-size energy can therefore be written in the form
\begin{equation}
E_{\mathrm{trial}}(\ell)
\simeq
-6JN
+AJ(N\ell)^{1/2}
+B(J+2\kappa)\frac{N}{\ell},
\label{eq:annex_crystal_energy}
\end{equation}
where \(A\) and \(B\) are geometric prefactors.

Defining the finite-size excess above the bulk contribution as $\Delta E(\ell)=E_{\mathrm{trial}}(\ell)+6JN$, minimizing Eq.~\eqref{eq:annex_crystal_energy} with respect to $\ell$ gives

\begin{equation}
\ell^\ast
\sim
N^{1/3}(1+2\mathcal{K})^{2/3},
\qquad
\frac{\Delta E(\ell^\ast)}{N}
\sim
J(1+2\mathcal{K})^{1/3}N^{-1/3}.
\label{eq:annex_crystal_scaling}
\end{equation}
Thus, for fixed \(\mathcal{K}\), the surface and bending
corrections vanish per monomer as \(N\rightarrow\infty\). This construction therefore approaches
\[
\frac{E_{\mathrm{trial}}}{JN}\rightarrow -6,
\]
providing the required admissible sequence for the crystalline
free-energy upper bound used in the main text. The construction is not
intended to select the microscopic crystalline morphology; any
alternative ordered compact structure with a lower free energy would
only strengthen the upper-bound argument.

\makeatletter
\def\bibsection{\section*{\refname}}
\makeatother
\bibliographystyle{apsrev4-2}
\bibliography{references}
\end{document}